\documentclass[floatfix,superscriptaddress,a4paper,
               showpacs,showkeys,nofootinbib]{revtex4}
\usepackage{epsfig}
\usepackage{latexsym}
\usepackage{xspace}
\usepackage{hyperref}
\usepackage[utf8]{inputenc}
\usepackage{indentfirst}
\usepackage{enumerate}
\usepackage{color}

\usepackage{amsmath}
\usepackage{amssymb}
\usepackage{array}
\usepackage[english]{babel}
\usepackage{url}
\newcommand{\eq}[1]{\begin{align} #1 \end{align}}

\begin{document}

\title{Hadron yields and fluctuations at the CERN Super Proton Synchrotron:\\
system size dependence from Pb+Pb to p+p collisions
}

\author{A. Motornenko}
\affiliation{Frankfurt Institute for Advanced Studies, Giersch Science Center, D-60438 Frankfurt am Main, Germany}
\affiliation{Institut f\"{u}r Theoretische Physik, Goethe Universit\"{a}t Frankfurt, D-60438 Frankfurt am Main, Germany}

\author{V.V. Begun}
\affiliation{Warsaw University of Technology, Faculty of Physics, Koszykowa 75, 00-662 Warsaw Poland}

\author{V. Vovchenko}
\affiliation{Frankfurt Institute for Advanced Studies, Giersch Science Center, D-60438 Frankfurt am Main, Germany}
\affiliation{Institut f\"{u}r Theoretische Physik, Goethe Universit\"{a}t Frankfurt, D-60438 Frankfurt am Main, Germany}

\author{M.I. Gorenstein}
\affiliation{Bogolyubov Institute for Theoretical Physics, 03680 Kiev, Ukraine}
\affiliation{Frankfurt Institute for Advanced Studies, Giersch Science Center, D-60438 Frankfurt am Main, Germany}

\author{H. Stoecker}
\affiliation{Frankfurt Institute for Advanced Studies, Giersch Science Center, D-60438 Frankfurt am Main, Germany}
\affiliation{Institut f\"{u}r Theoretische Physik, Goethe Universit\"{a}t Frankfurt, D-60438 Frankfurt am Main, Germany}
\affiliation{GSI Helmholtzzentrum f\"ur Schwerionenforschung GmbH, D-64291 Darmstadt, Germany}

\begin{abstract}
The kaon to pion ratio $K^+/\pi^+$ and the scaled variance $\omega^-$ for fluctuations of negatively charged particles are studied within the statistical hadron resonance gas (HRG) model and the Ultra relativistic Quantum Molecular Dynamics (UrQMD) transport model.
The calculations are done for p+p, Be+Be, Ar+Sc, and Pb+Pb collisions at the CERN Super Proton Synchrotron energy range to reveal the system size dependence of hadron production. For the HRG calculations the canonical ensemble is imposed for all conserved charges.
In the UrQMD simulations the centrality selection in nucleus-nucleus collisions is done by calculating the forward energy $E_{\rm F}$ deposited in the Projectile Spectator Detector, and the acceptance maps of the NA61/SHINE detectors are used.
A comparison of the HRG and UrQMD results with  the data of the  NA61/SHINE Collaboration is done.
To understand a difference of the event-by-event fluctuations in p+p and heavy ion collisions the centrality selection procedure in the sample of all inelastic p+p events is proposed and analyzed within the UrQMD simulations.
\end{abstract}



\date{\today}

\maketitle

\section{Introduction}\label{Sec-Intro}
A main goal of the experimental programs in high energy nucleus-nucleus (A+A) collisions is to form and study a new state of matter -- the quark gluon plasma (QGP). Proton-proton (p+p) reactions at the same energy per nucleon serve then as the `reference point', e.g., the strangeness enhancement \cite{Rafelski:1982pu} and $J/\psi$ suppression \cite{Matsui:1986dk} observed in central heavy ion collisions have been often referred as  the QGP signatures.

An ambitious experimental program for a search of the critical point related to the deconfinement transition is carried out by the NA61/SHINE Collaboration at the Super Proton Synchrotron (SPS) of the European Organization for Nuclear Research (CERN)~\cite{Gazdzicki:995681,Abgrall:2014xwa,Aduszkiewicz:2642286}.
The program includes a variation in the atomic mass number $A$ of the colliding nuclei as well as an energy scan.
Using these data one hopes to scan  the phase diagram in the plane of temperature $T$ and baryon chemical potential $\mu_B$ and locate a position of the critical point by studying its event-by-event (e-by-e) fluctuation signals.
The results on p+p, Be+Be, and Ar+Sc collisions have been presented by the NA61/SHINE Collaboration \cite{Abgrall:2014xwa,Gazdzicki:995681,Aduszkiewicz:2287091},
while the Pb+Pb data were presented earlier by the NA49 Collaboration \cite{Alt:2008qm,Alt:2008iv,Alt:2004kq,Anticic:2011ny,Friese:2002re,NA49_compil}.
The studied collisions and the projectile momenta are the following:
\begin{table}[h!]
\centering
\setlength{\tabcolsep}{0.5em}
\begin{tabular}{l|>{\centering\arraybackslash}p{1cm}|>{\centering\arraybackslash}p{1cm}|>{\centering\arraybackslash}p{1cm}|>{\centering\arraybackslash}p{1cm}|>{\centering\arraybackslash}p{1cm}|>{\centering\arraybackslash}p{1cm}}
&\multicolumn{6}{c}{$p_{\rm lab}~(A\,{\rm GeV}/c)$}\\ \hline
NA49: Pb+Pb & $-$ & 20 & 30 & 40 & 80 & 158 \\ \hline
NA61: p+p & 13 & 20 & 31 & 40 & 80 & 158 \\ \hline
NA61: Be+Be and Ar+Sc & 13 & 19 & 30 & 40 & 75 & 150 \\
\end{tabular}
\end{table}

Centrality selection of collision events is important for a proper physical interpretation of the measured fluctuations. In peripheral collisions of heavy ions one can not avoid strong fluctuations of the number of nucleon participants. These participant number fluctuations become the dominant ones in
all e-by-e fluctuations of final hadrons (see a discussion of this point in Refs.~\cite{Konchakovski:2007ss, Konchakovski:2007ah,Skokov:2012ds,Begun:2016sop,Braun-Munzinger:2016yjz,Broniowski:2017tjq,Motornenko:2017klp}).
The preliminary data on Be+Be collisions from the NA61/SHINE Collaboration provoked speculations on `the onset of fireball' -- rather specific change of hadron production properties when moving from small systems, like those created in p+p and light nuclei (Be+Be) to central collisions of heavy nuclei like Ar+Sc and Pb+Pb in the SPS energy region \cite{Aduszkiewicz:2287091,Seryakov:2017jpr,Gazdzicki:2017zrq}.
The two observables were presented: a ratio of the mean hadron multiplicities $\langle K^+\rangle/\langle \pi^+\rangle$ and the scaled variance for negatively charged particles
\eq{\label{om}
\omega[N_-]~\equiv~\frac{\langle N_-^2\rangle~-~\langle N_-\rangle^2}{\langle N_-\rangle}~\equiv~\omega^-,
}
where $\langle \ldots \rangle$ denotes the e-by-e  averaging\footnote{The kaon to pion ratio
measured at the central rapidities will be denoted as $K^+/\pi^+$.}.

At the top SPS energy 150~$A$ (158 $A$)\,GeV/$c$ the experimental value of the scaled variance $\omega^-$ in p+p collisions appears to be larger than in central Ar+Sc and Pb+Pb collisions. On the other hand, the preliminary data in the central Be+Be collisions correspond to the value of $\omega^-$ even larger than that in the p+p reactions. The $K^+ / \pi^+$ ratio at the SPS energies remain approximately the same for p+p and Be+Be collisions, but this ratio becomes (much) larger for heavy ions  like Ar+Sc and Pb+Pb. The above results admit a possibility of the step-like behavior of $K^+ / \pi^+$ and $\omega^-$  as functions of the system size with almost no changes between p+p and Be+Be, and a significant change between Be+Be and Ar+Sc. Thus, the collisions of light nuclei seem to look as the samples of independent nucleon-nucleon collisions, while the central collisions of heavy ions lead to the fireball formation. A similar physical picture was suggested in terms of the parton percolation model \cite{Satz:1998kg}. What is an exact value of the atomic number $A$ of colliding nuclei where the `onset of fireball'  takes place? How rapid is the transition from  independent nucleon-nucleon collisions observed in reactions with the light nuclei to a fireball-like behavior seen in the central A+A collisions  of heavy nuclei? The answers to these questions are still unclear.

A goal of this paper is to probe the effects observed by the NA61/SHINE Collaboration by the statistical hadron resonance gas (HRG) and the transport Ultra relativistic Quantum Molecular Dynamics (UrQMD) \cite{Bass:1998ca, Bleicher:1999xi} models. These two approaches are both popular and successful phenomenological models to describe the hadron production in A+A collisions. We use both models to calculate $K^+/\pi^+$ and $\omega^-$ in p+p, Be+Be, Ar+Sc, and Pb+Pb collisions at the SPS energy region.

The paper is organized as follows. Section~\ref{Sec-HRG} presents the canonical ensemble (CE) HRG model calculations. The results of the UrQMD simulations are presented in Sec.~\ref{Sec-UrQMD}. Sections \ref{Sec-HRG} and \ref{Sec-UrQMD} include also a comparison of both models  with the available data. Section~\ref{Sec-Summary} summarizes the paper.

%
\section{$K^+/\pi^+$ ratio and $\omega^-$ in the CE HRG }\label{Sec-HRG}
%
%
In the central collisions of heavy nuclei a formation of the statistical system (fireball) is expected. The HRG model is used for a description of the latest stage of the fireball evolution, the chemical freeze-out. There are also numerous successful attempts to use the HRG model for a description of the hadron yields produced in p+p reactions, see, e.g., \cite{Becattini:1997rv, Vovchenko:2015idt,Chatterjee:2016cog,Begun:2018qkw}, and references therein. In this case, due to a small size of the system an exact charge conservation for each microstate of the statistical system becomes important. Thus, the CE should be used. In the present paper we use the CE HRG calculations for all systems in order to make a systematic comparison of the hadron production for both small and large systems within the HRG model. For the hadron yields in Pb+Pb collisions at the SPS energies most of the CE results for hadron yields converge to the grand canonical ensemble (GCE) results. This is known as the thermodynamic equivalence of the statistical ensembles. In the GCE, only the average values of the conserved charges are fixed, but their changes from one microstate to another are allowed. Therefore, the e-by-e hadron number fluctuations are essentially different in the CE and GCE even in the thermodynamic limit \cite{Begun:2004zb}.

\subsection{Canonical ensemble results}
We use the CE HRG for non-interacting hadrons and resonances, presented in details in the recent papers \cite{Vovchenko:2015idt,Begun:2018qkw}. The calculations were done using \texttt{Thermal-FIST} package~\cite{ThermalFIST}, which is publicly available. In the CE the following parameters are introduced
 \eq{\label{par}
 T,~V,~\gamma_S\,,
 }
where $T$ and $V$ are, respectively, the system temperature and volume at the chemical freeze-out, and $\gamma_S$~\cite{Letessier:1998sz,Letessier:2005qe,Rafelski:2015cxa} is the strangeness suppression parameter, which takes into account incomplete chemical equilibration of strange hadrons. In what follows we use the system radius $R\equiv [3V/(4\pi)]^{1/3}$ instead of volume $V$ for convenience. The baryon number $B$, electric charge $Q$ and net strangeness $S=0$ of the considered system are strictly fixed by the corresponding values of colliding nuclei.

\begin{figure}[t!]
\includegraphics[width=0.49\textwidth]{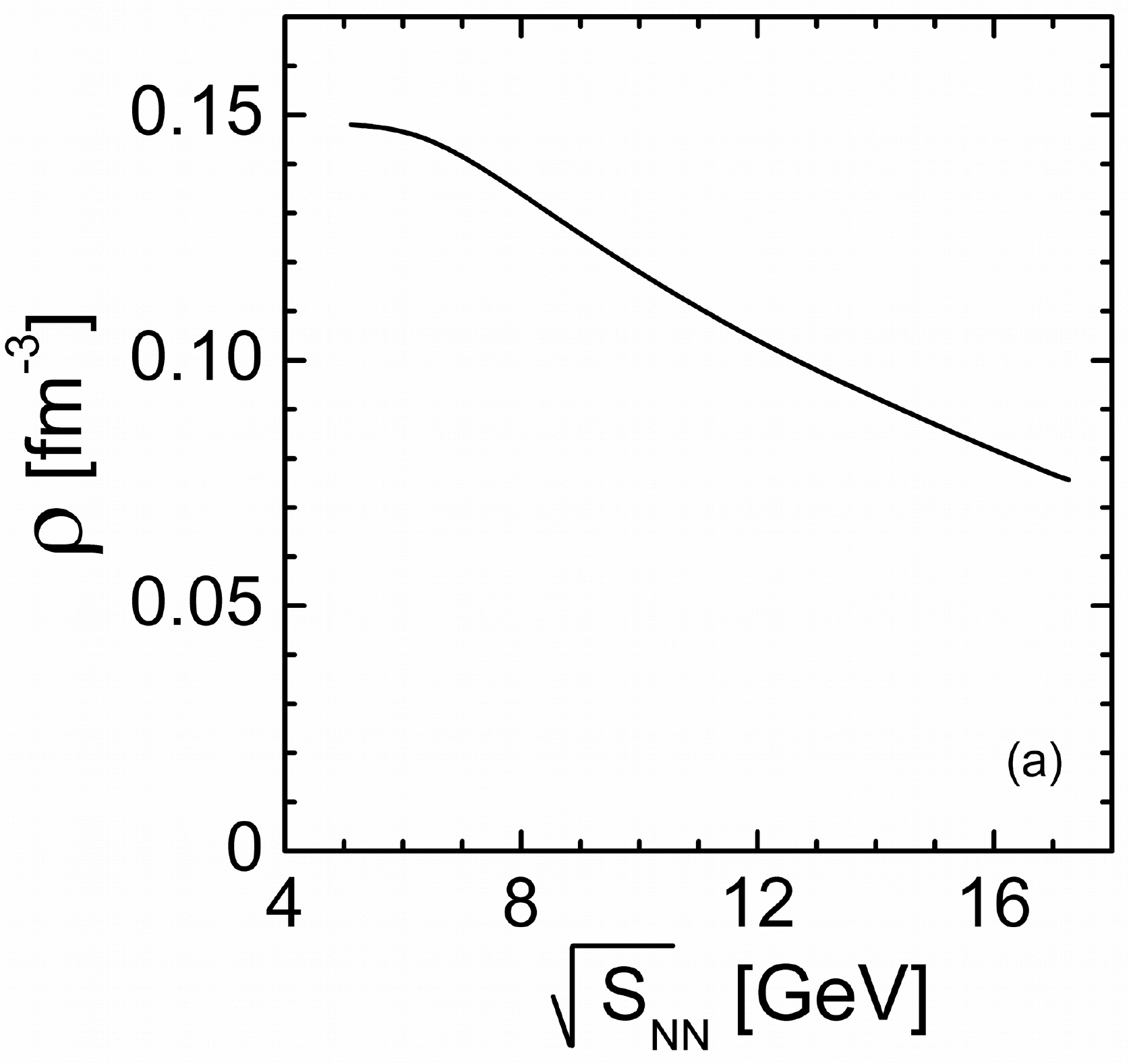}
\includegraphics[width=0.49\textwidth]{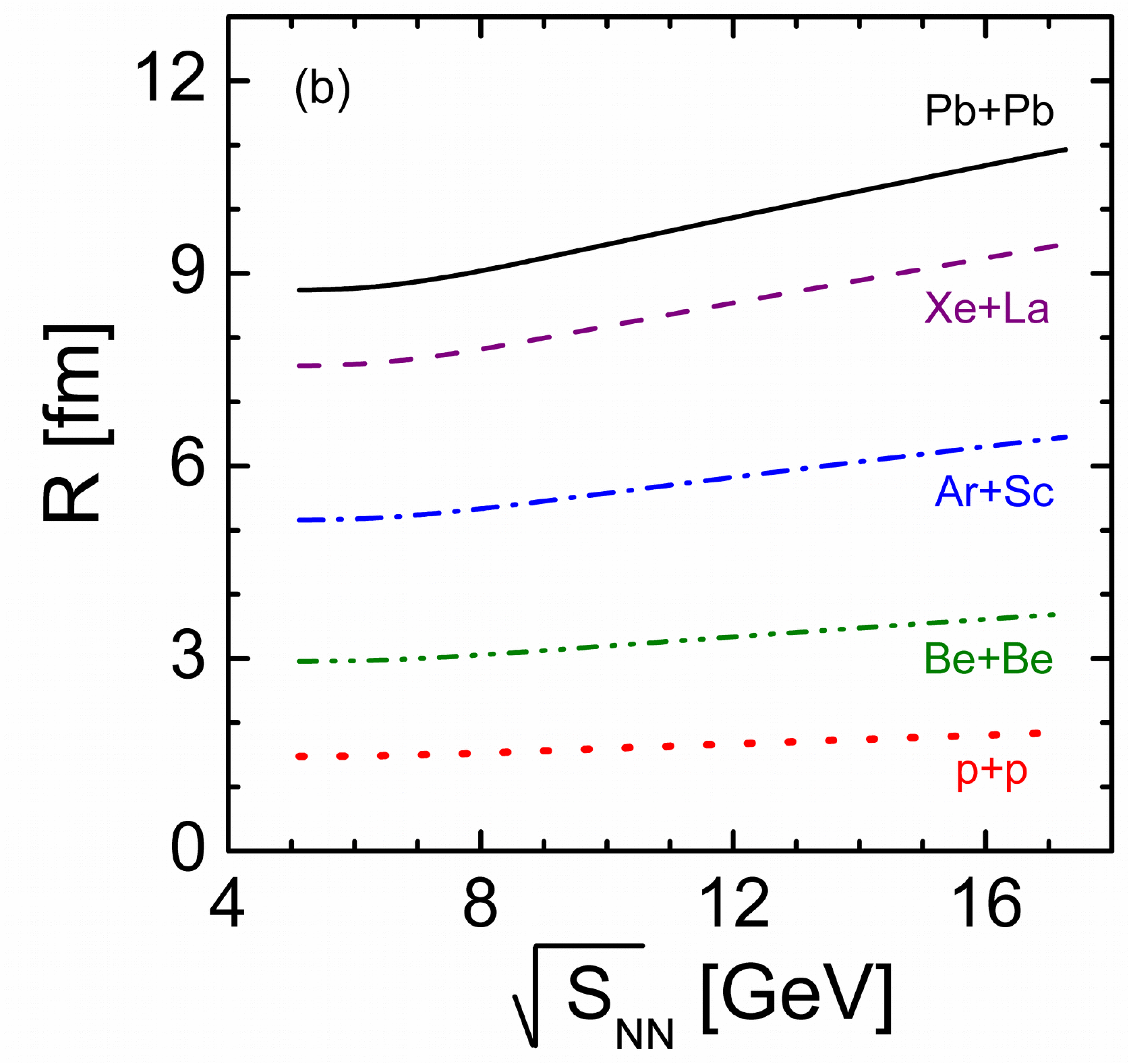}
\caption{The baryon density $\rho_B$ (a) and the radius $R$ (b) of the statistical system at the chemical freeze-out as functions of the collision energy.
}
\label{fig-1}
\end{figure}

To clarify the system size dependence in most simple and transparent way, we assume first a complete strangeness equilibrium, $\gamma_S=1$, and the same value of temperature $T$ for all colliding systems. The values of $T$ at different collision energies are taken in the following form \cite{Cleymans:2005xv}:
\eq{\label{T}
T(\mu_B)~=~a-b\,\mu_B^2-c\,\mu_B^4~,&& \mu_B(\sqrt{s_{NN}})~=~ \frac{d}{1~+~e\,\sqrt{s_{NN}}}
}
where the parameters $a$, $b$, $c$, $d$, and $e$ are fixed from the fit to the central Pb+Pb data at the SPS energies in Ref.~\cite{Vovchenko:2015idt}. The $\sqrt{s_{NN}}$ is the center of mass energy of the nucleon pair in the colliding systems. Therefore, only the size $R$ of the considered system at the chemical freeze-out remains different for different colliding nuclei.
To find the system size at the freeze-out for different colliding nuclei we use an approximation of equal baryon densities $\rho_B$ in all considered A+A systems, and assume that this value of $\rho_B$ is equal to the values found in the GCE HRG with Eq.~(\ref{T}). The function $\rho_B(\sqrt{s_{NN}})$ calculated in this way is shown in Fig.~\ref{fig-1} (a). The obtained radius $R=[3B/(4\pi \rho_B)]^{1/3}$, where total baryonic number $B$ equals to the number of nucleon participants, is shown in Fig.~\ref{fig-1} (b)  for the systems studied by the NA61/SHINE Collaboration. Note that the numerical values of $\rho$ and $R$ should be interpreted with care because of the excluded volume corrections effects neglected in the present paper. A presence of nonzero baryon proper volumes would  lead to smaller  $\rho_B$ and larger $R$ values \cite{Vovchenko:2018cnf}.

\begin{figure}[h!]
\includegraphics[width=0.49\textwidth]{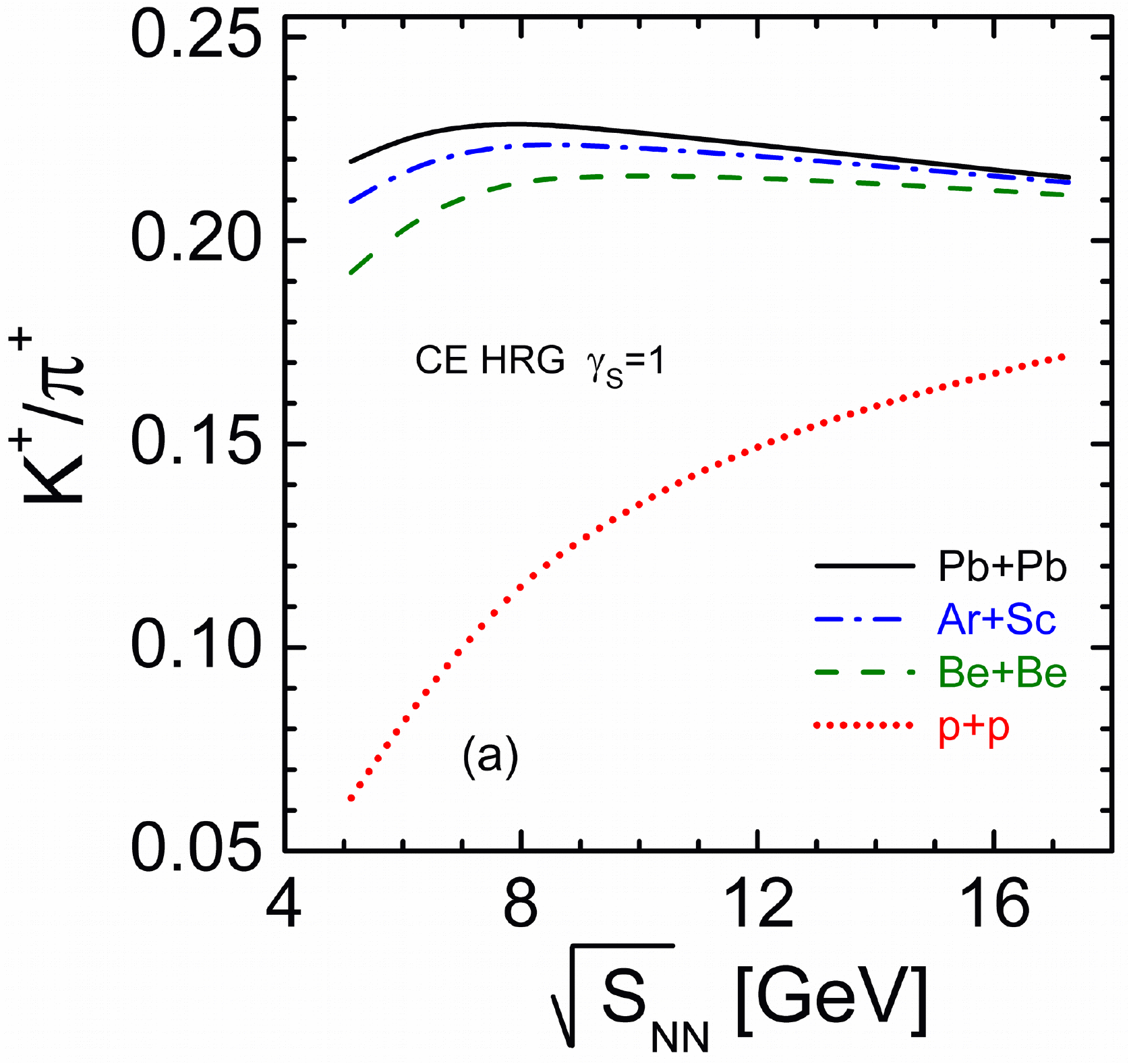}~~~
\includegraphics[width=0.49\textwidth]{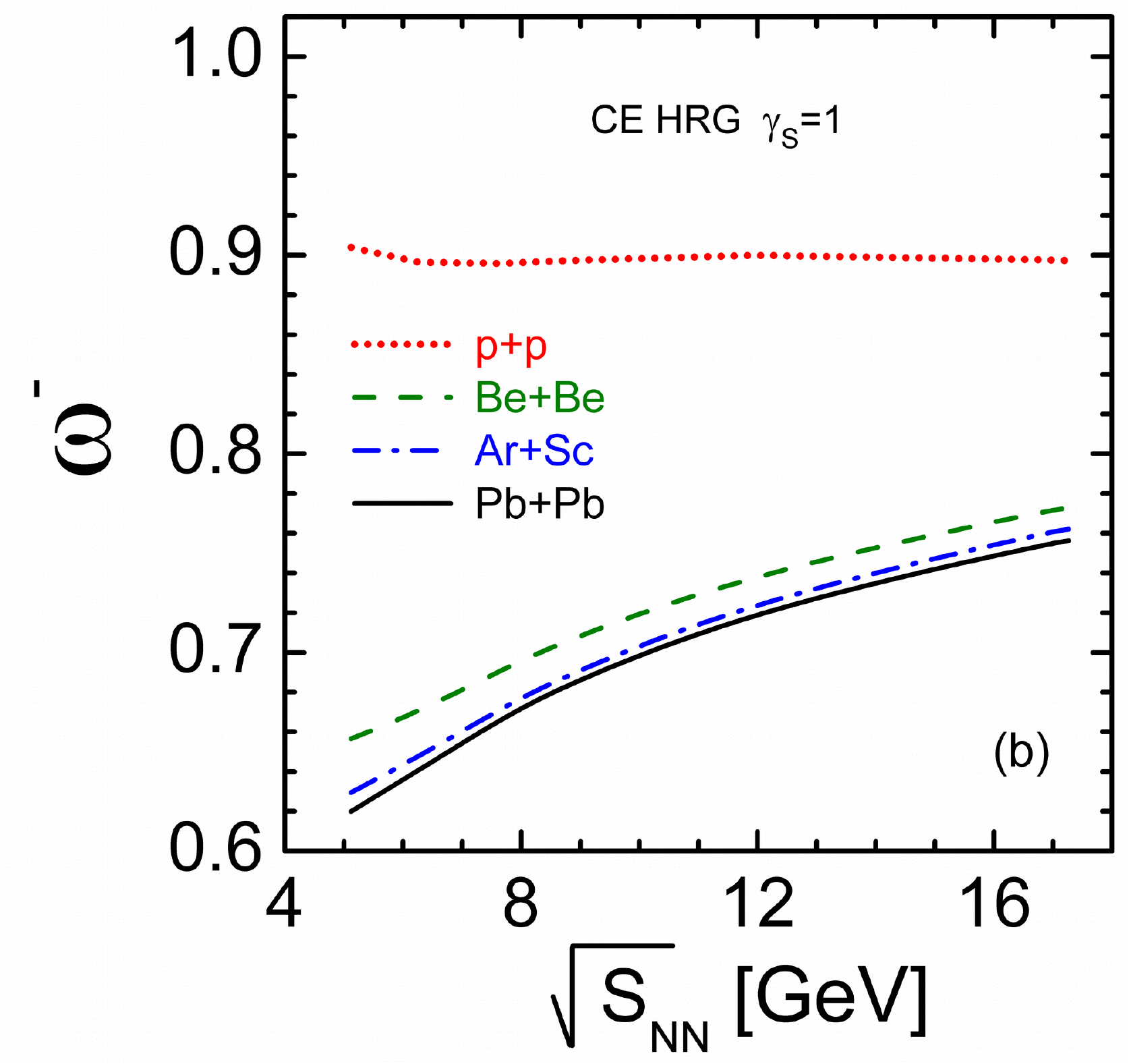}
\caption{The results of the CE HRG calculations for the $K^+ / \pi^+$ ratio (a) and the scaled variance $\omega^-$ of negatively charged particles (b) as functions of the collision energy.
}
\label{fig-2}
\end{figure}

The corresponding values of $K^+ / \pi^+ $ and $\omega^-$ calculated in the CE HRG for non-interacting hadron species are shown in Figs.~\ref{fig-2} (a) and (b), respectively, as functions of the collision energy.

Within HRG the intensive quantities, particularly $K^+/\pi^+$ and $\omega^-$, are the functions of $T$ and $\mu_B$, and they are not sensitive to the system volume $V$ in the GCE. This is not the case for the CE considered in our studies. The global charge conservation suppresses the mean multiplicities of hadrons in small systems. Significant CE suppression effects happen for hadron yields when the total number of particles and antiparticles of that corresponding conserved charge is of an order of unity or smaller. This leads, e.g., to $\langle K^+\rangle_{\rm CE}< \langle K^+\rangle_{\rm GCE}$ at finite values of $V$ because of the exact strangeness conservation. This difference of the hadron yields in the CE and GCE becomes negligible in the thermodynamic limit $V\rightarrow \infty$. On the other hand, the smaller CE values of the particle number fluctuations, e.g. $\omega^-_{\rm CE}< \omega^-_{\rm GCE}\cong 1$, are just most pronounced in the thermodynamic limit.

From Fig.~\ref{fig-2} one can see the CE effects due to the charge conservation, i.e., $\left(K^+ / \pi^+ \right)_{\rm A+A}> \left(K^+ / \pi^+ \right)_{\rm p+p}$ and $\left(\omega^-\right)_{\rm A+A} < \left(\omega^-\right)_{\rm p+p}$. Even more, the $K^+/\pi^+$ ratio increases and $\omega^-$ decreases monotonously with the size of colliding systems. Therefore, the CE suppression is stronger for the particle {\it yields}  in {\it small systems} like p+p, and for particle number {\it fluctuations} in {\it large systems} like Pb+Pb, Figs.~\ref{fig-2} (a) and  (b), respectively  (see also Ref.~\cite{Begun:2004zb}).
Note that the differences between p+p and A+A collisions are stronger at the smallest collision energies. These features of the CE HRG are similar to those observed in the data.
However, contrary to the data, practically the whole system size dependence in the CE HRG model occurs in between p+p and Be+Be, and a little change is seen in between Be+Be and Pb+Pb.
%

%
\subsection{Incomplete strangeness equilibration and acceptance corrections in HRG}
A quantitative comparison of the CE HRG model results with the data
requires several further steps. An incomplete equilibration of the strange hadrons
and a finite detector acceptance
should be taken into account.

\begin{figure}[h!]
\includegraphics[width=0.49\textwidth]{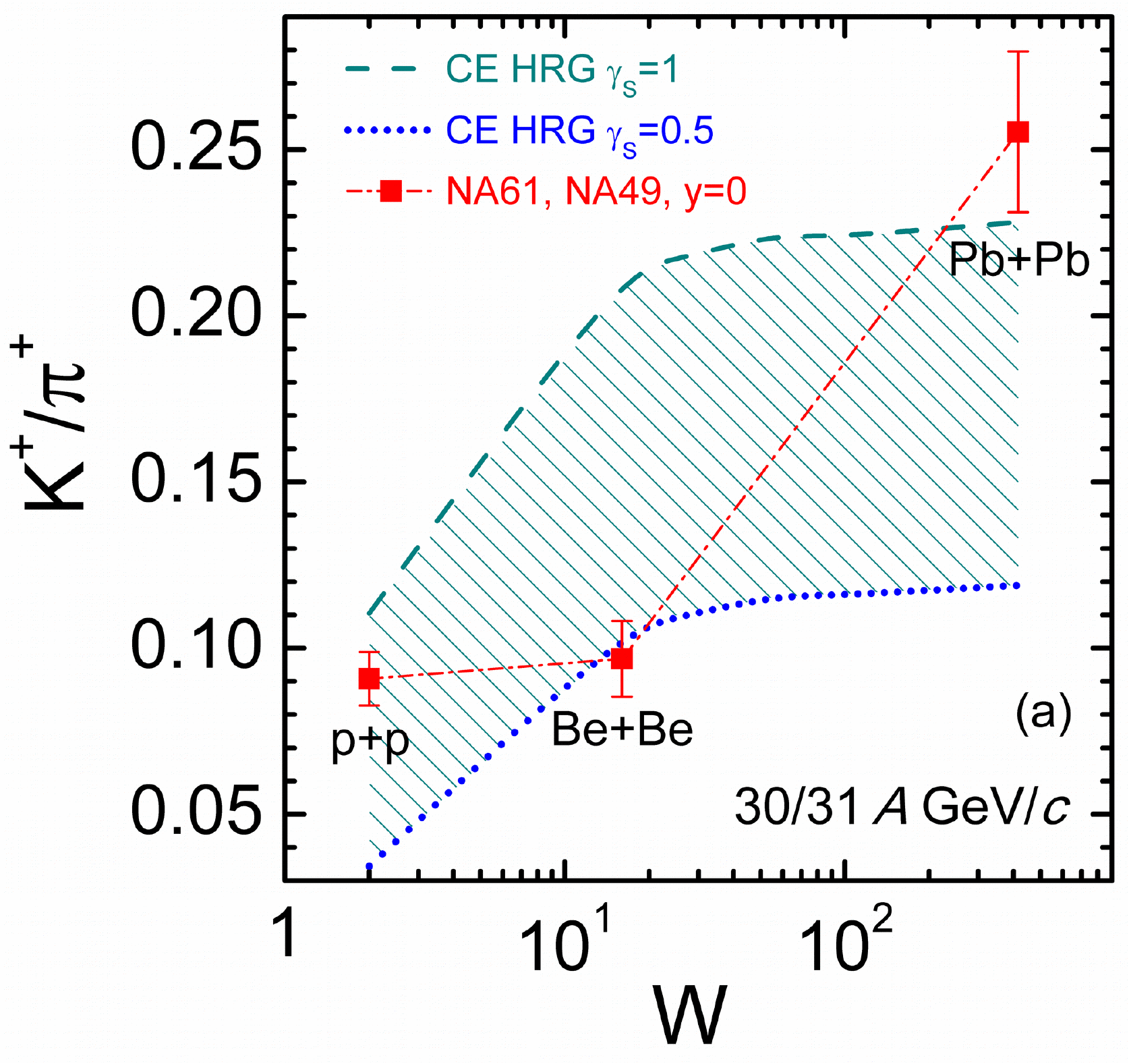}~~
\includegraphics[width=0.49\textwidth]{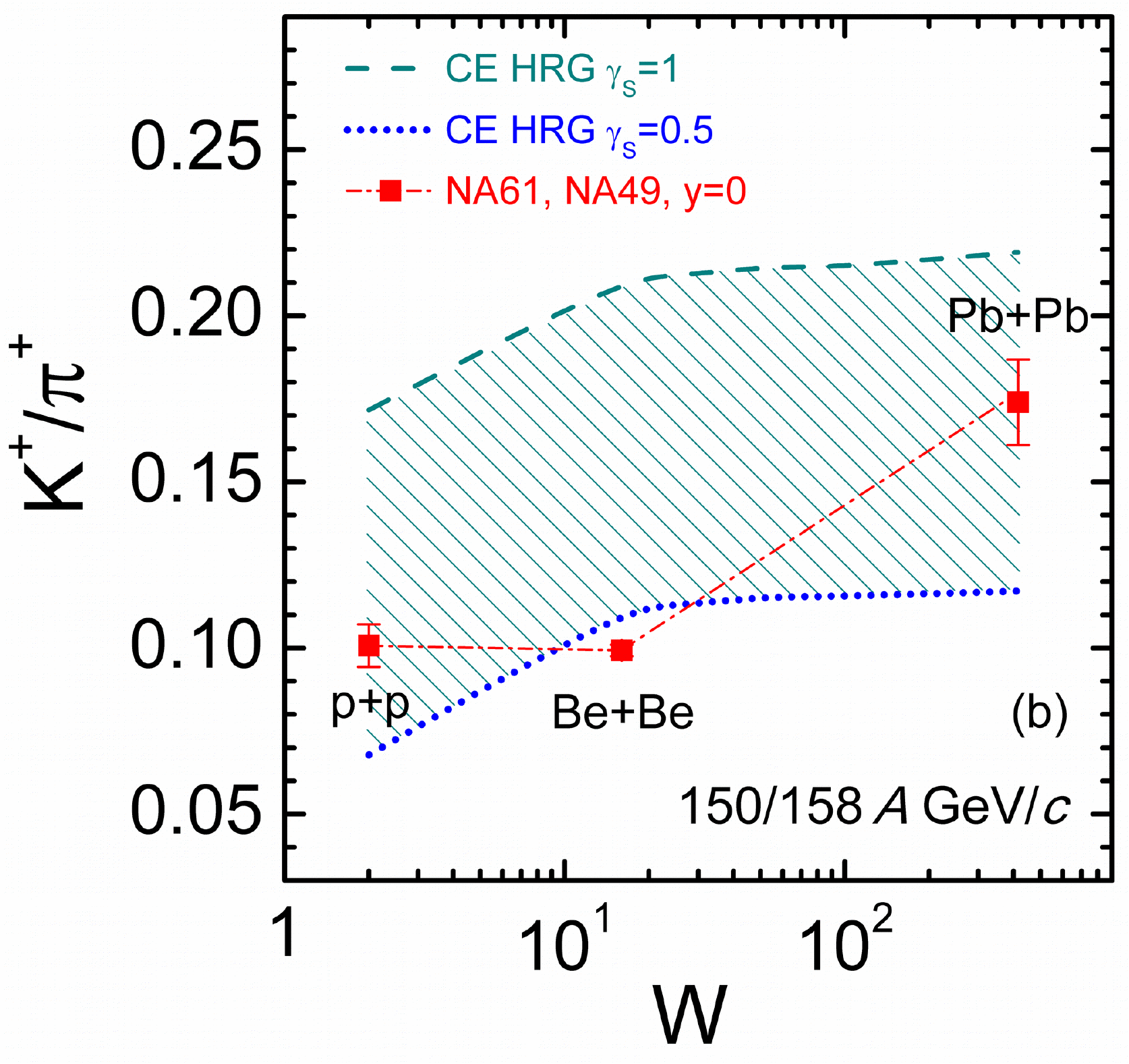}
\\[0.2cm]
\includegraphics[width=0.49\textwidth]{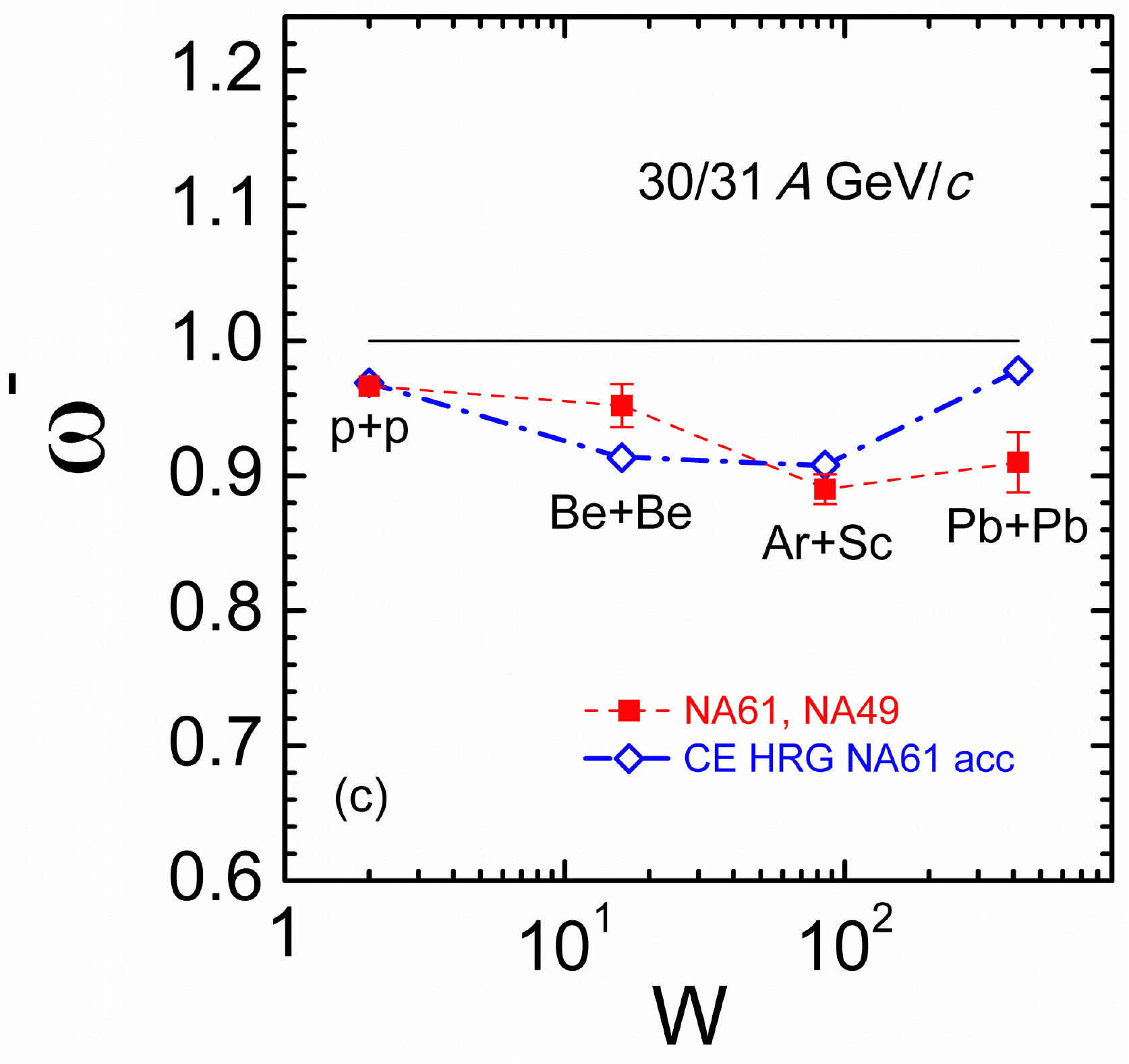}~~
\includegraphics[width=0.49\textwidth]{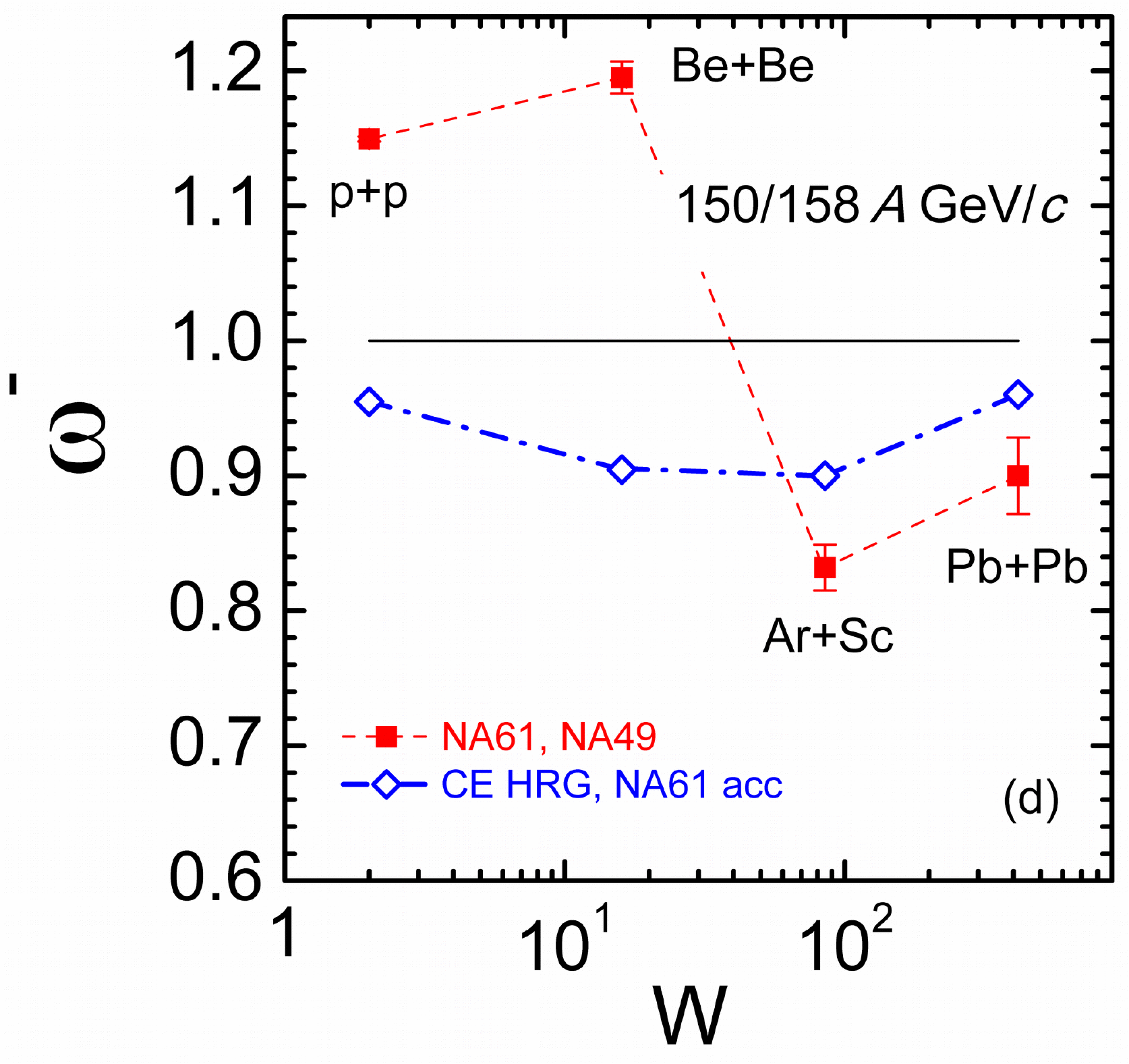}
\caption{The CE HRG results for $K^+/\pi^+$  (a-b) and  $\omega^-$ (c-d) as functions of $W\equiv A_1+A_2$ which equals to the total number of nucleons in the colliding nuclei. See text for details. The data are shown by  red squares for $ K^+/\pi^+$ ~\cite{Aduszkiewicz:2017sei,Aduszkiewicz:2017mei,Alt:2008qm} and for $\omega^-$~\cite{Seryakov:2017jpr,Alt:2007jq}.
}
\label{fig-3}
\end{figure}

Many previous fits of the hadron multiplicity data within the HRG model, both in the GCE and CE, demonstrated an incomplete chemical equilibration of the strange hadrons. This is usually compensated by the strangeness suppression parameter $\gamma_S$ \cite{Letessier:1993hi,Letessier:1998sz,Letessier:2005qe,Rafelski:2015cxa} with numerical values in the range of $0.5<\gamma_S<1$ for the SPS energies. The number of strange and antistrange particles becomes then smaller than that expected in the equilibrium HRG. The suppression factor $\gamma_S^n$ for each hadron depends on the number $n$ of strange quarks and/or antiquarks in a given hadron.
The small values of $\gamma_S\sim 0.5$ correspond to p+p and small SPS energies, larger $\gamma_S\sim 1$ to central collisions of heavy nuclei and large SPS energies~\cite{Castorina:2016eyx}.
A comparison of the CE HRG results with the available data are presented in Fig.~\ref{fig-3}. Dashed and dotted lines in Figs.~\ref{fig-3} (a) and (b) present the CE HRG results with,  $\gamma_S=1$ and $\gamma_S=0.5$, respectively.

The HRG model calculates the hadron multiplicities in the whole phase-space, the so-called full $4\pi$ acceptance. However, e-by-e hadron yield measurements are done within the finite acceptance of the detectors.
In order to correct the HRG scaled variance with respect to the experimental acceptance the following simple formula is often used (see, e.g., \cite{Begun:2004zb})
\eq{
\label{acc}
\omega^-~=~1~-q~+q\,\omega^-_{4\pi} ~.
}
In Eq.~(\ref{acc}),  $q=\langle N_-\rangle/\langle N_-\rangle_{4\pi}$, where $\langle N_-\rangle$ and $\langle N_-\rangle_{4\pi}$ are the average $N_-$ values  in the accepted region and in the full phase space, respectively, and  $\omega^-$ and $(\omega^-)_{4\pi}$ denote the scaled variances of the accepted hadrons and all final hadrons, respectively. Equation (\ref{acc}) assumes the binomial acceptance probability which, however, can be violated (see, e.g.,  Ref.~\cite{Motornenko:2017klp}).
The CE HRG results presented in Figs.~\ref{fig-3} (c) and (d) are calculated for $\gamma_S=1$. The value of $\omega^-$ is almost independent of $\gamma_S$, as the majority of negatively charged particles are $\pi^-$ mesons. The CE HRG predicts a monotonous decrease of the 4$\pi$ values of $\omega^-$ with atomic number of colliding nuclei. This is seen in Fig.~\ref{fig-2} (b). The results presented in Figs.~\ref{fig-3} (c) and (d) are based on  Eq.~(\ref{acc}) with the acceptance parameter $q$  estimated by the Monte-Carlo simulations of the detector acceptance. The effects of a non-monotonous behavior of $\omega^-$ with $A$ seen in Figs.~\ref{fig-3} (c) and (d) for the accepted particles  appear only because of the smaller values of the acceptance parameter $q$ in Eq.~(\ref{acc}) for Pb+Pb collisions ($q\cong 0.06$ and $q\cong 0.16$ for 30 and 158~$A\,$GeV$/c$, respectively) in  comparison to other A+A and p+p reactions ($q\cong 0.3$ and $q\cong 0.4$ for 30/31 and 150/158~$A\,$GeV$/c$, respectively).

The value of $\omega^- <1$ are observed for all colliding systems at 30/31~$A$~GeV/$c$ and for large systems, namely Ar+Sc and Pb+Pb, at the highest available SPS energy, 150/158~$A$~GeV$/c$. All these data can be qualitatively described by the CE HRG. However, for the two smallest systems, p+p and Be+Be, at 150/158~$A$~GeV/$c$ the data correspond to the large values, $\omega^- > 1$. In these two cases the CE HRG results are in contradiction with the data.

%
\section{$K^+/\pi^+$ ratio and $\omega^-$ in the UrQMD model}
\label{Sec-UrQMD}

\subsection{Centrality Selection in A+A collisions}

The NA61/SHINE Collaboration uses the Projectile Spectator Detector (PSD) to measure the forward energy $E_{\rm F}$. This is the energy deposited in a small angle of the forward hemisphere. The calorimeter is not able to identify particles, thus, not only the projectile spectators contribute to $E_{\rm F}$.
To make a proper comparison with the data the experimental centrality selection and acceptance are imposed in our UrQMD simulations. This is done in Be+Be and Ar+Sc collisions with the maps of the PSD and NA61/SHINE detectors~\cite{PSD-maps}. In Pb+Pb collisions the zero value of the impact parameter $b$ is used in the UrQMD simulations, and only the acceptance maps of the NA61/SHINE detectors are imposed. Pb+Pb collisions with $b=0$ correspond approximately to the 1\% most central events (see, e.g., Refs.~\cite{Alt:2007jq, Konchakovski:2007ah}).

\begin{figure}[h!]
\centering
\includegraphics[width=0.49\textwidth]{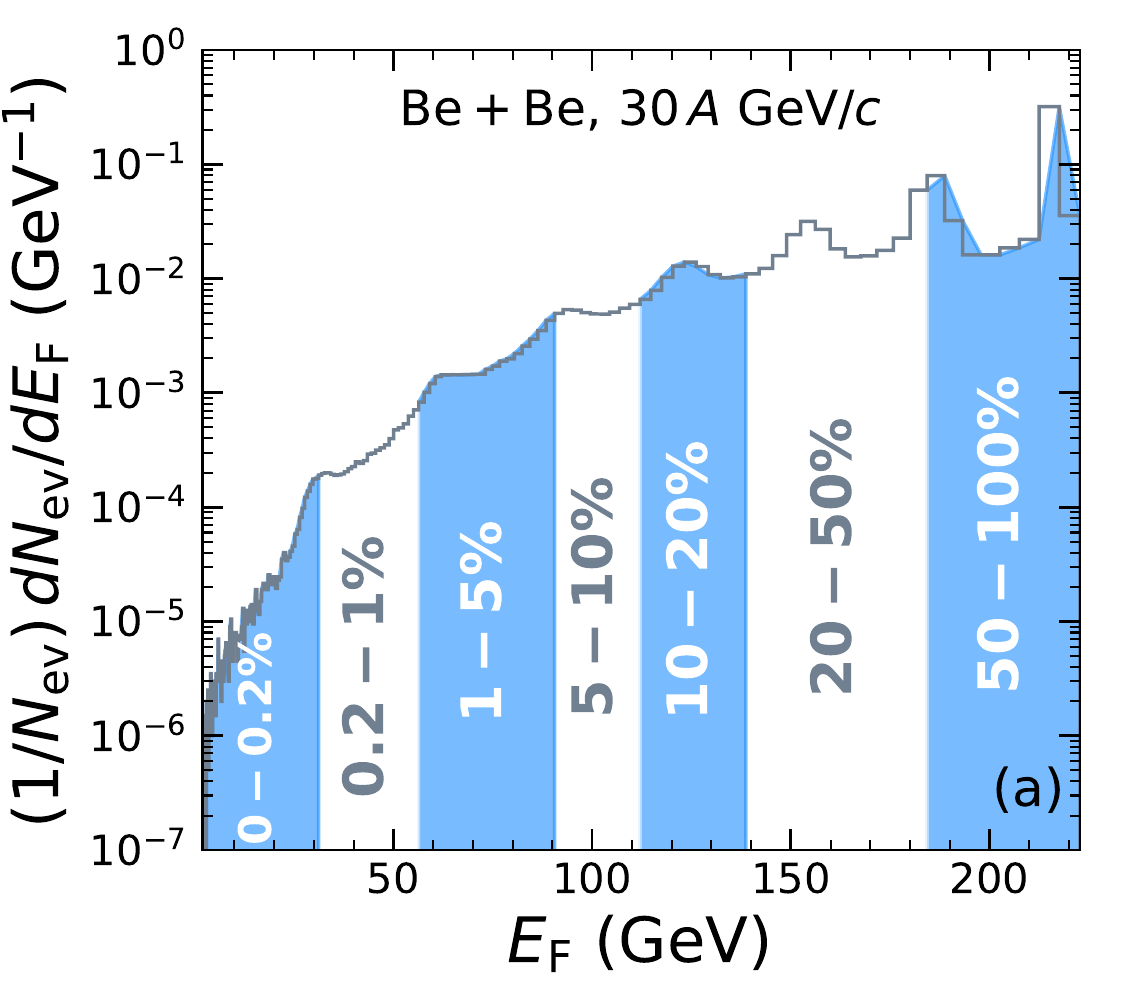}
\includegraphics[width=0.49\textwidth]{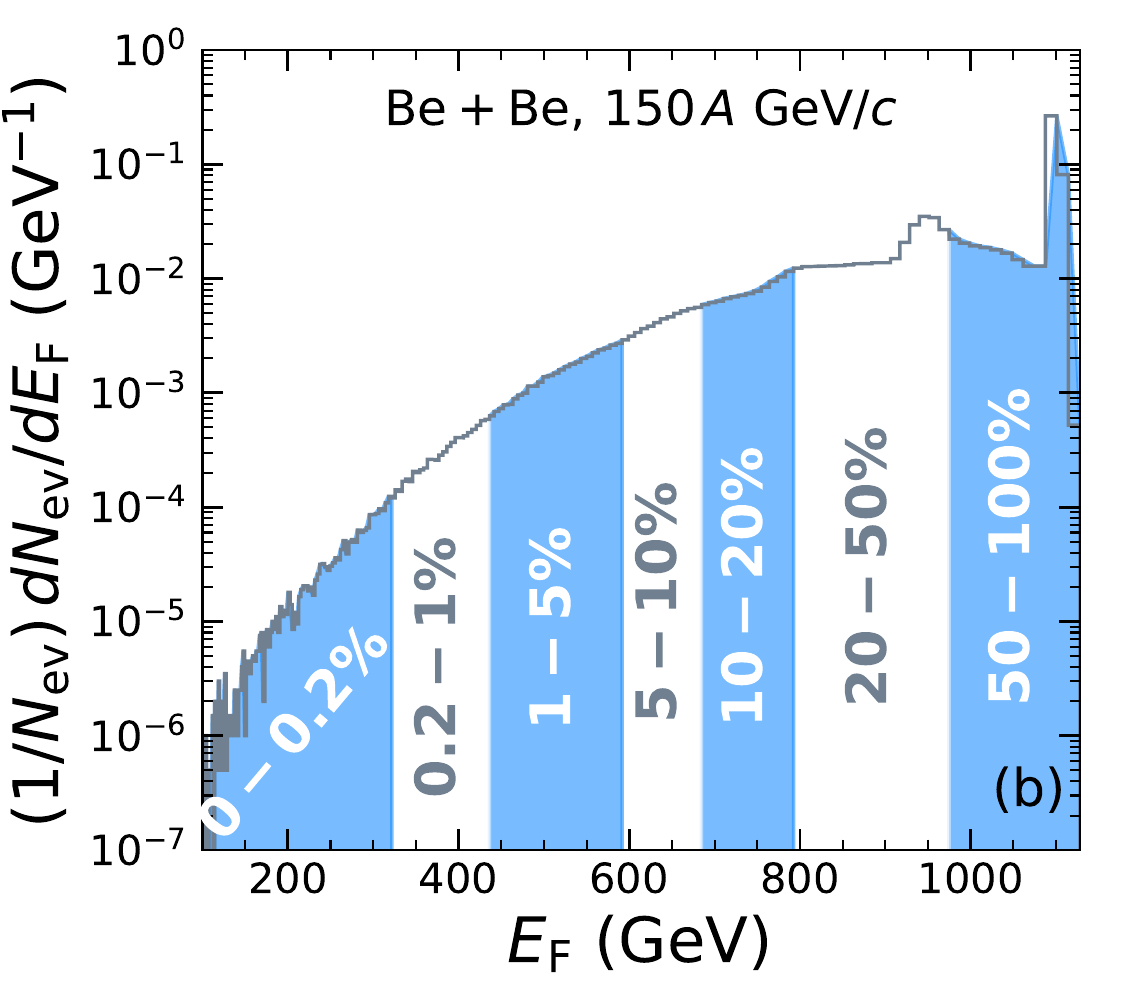}\\
\includegraphics[width=0.49\textwidth]{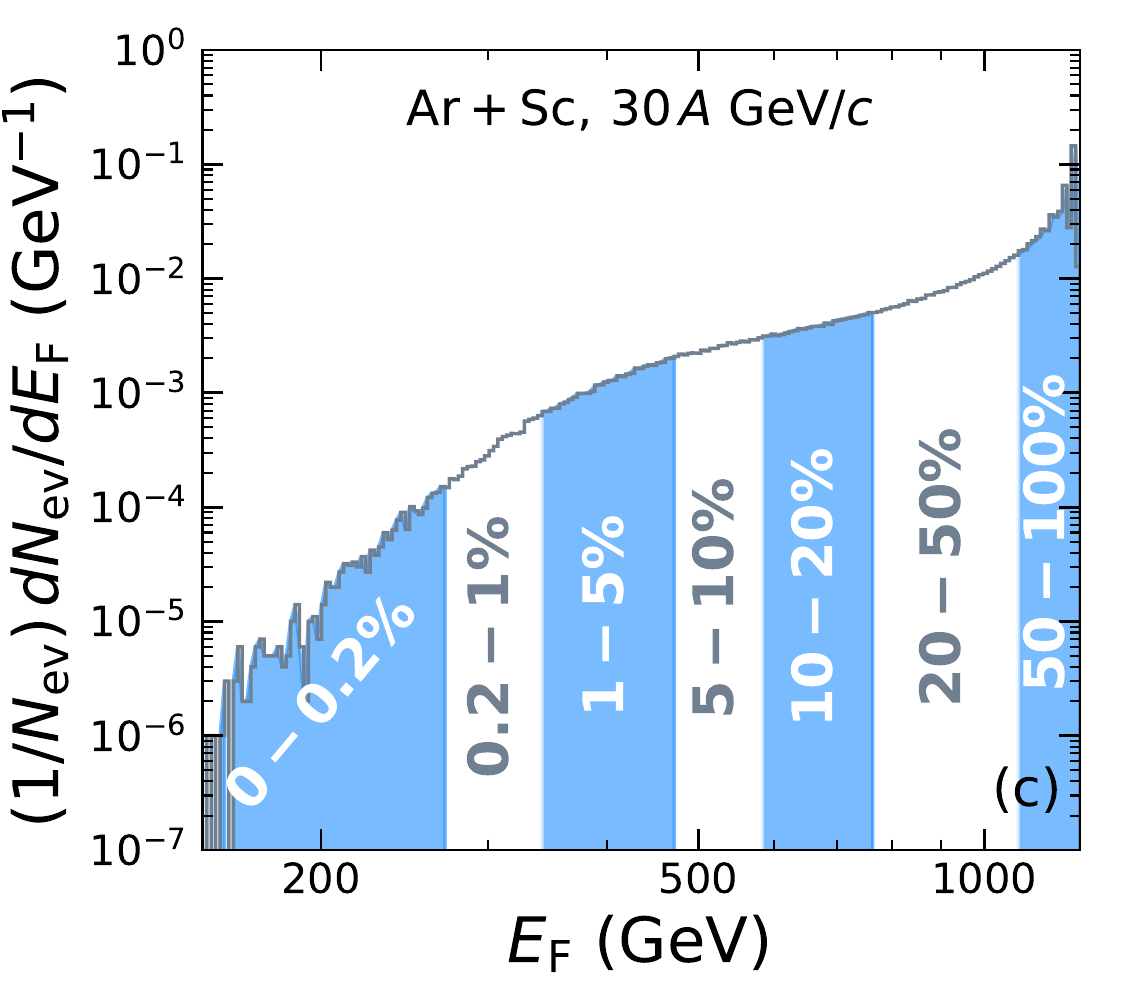}
\includegraphics[width=0.49\textwidth]{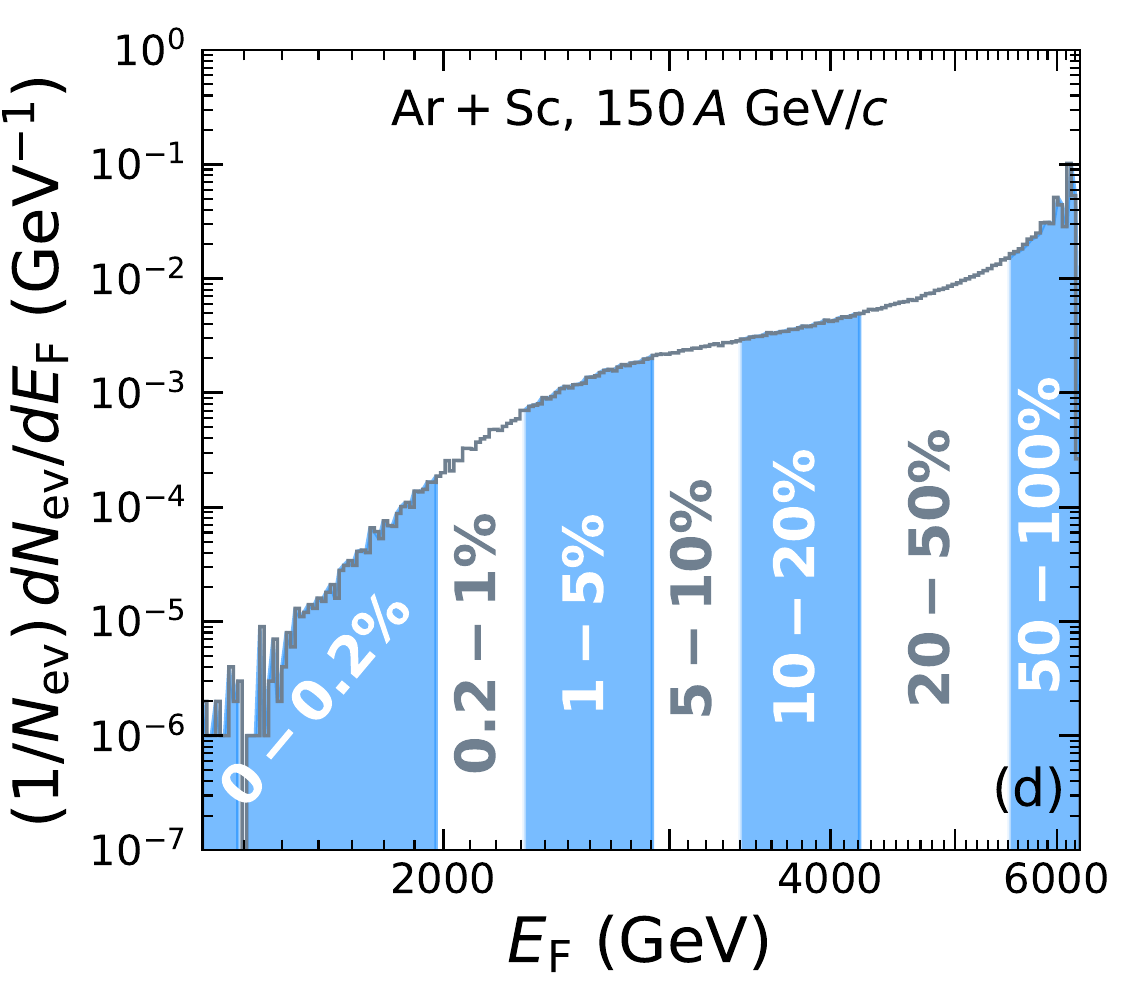}
\caption{Number of events as a function of the forward energy $E_{\rm F}$ calculated in the minimum bias UrQMD simulations. The colored regions correspond to different centrality classes that are indicated with respective numbers.}
\label{fig-Efwd}
\end{figure}

A selection of an appropriate centrality class is a special task. One needs a sample with the sufficient number of events. On the other hand, this sample can not be too wide to minimize the `background fluctuations`, e.g., geometrical fluctuations of the impact parameter $b$. The centrality classes obtained in the UrQMD simulations are presented in Fig.~\ref{fig-Efwd}. The event distributions by the $E_{\rm F}$ look rather monotonous where contributions from every single participant are smeared out by secondary particles that fly into forward calorimeter. However, for small systems and energies one is still able to observe peaks in the distributions that allow to identify contribution from a single participant. For example, in Be+Be at 30$A$~GeV one can observe 7 peaks that are produced by 1-7 spectator nucleons in the Be+Be collision, see  Fig.~\ref{fig-Efwd}(a).

\begin{figure}[h!]
\centering
\includegraphics[width=0.49\textwidth]{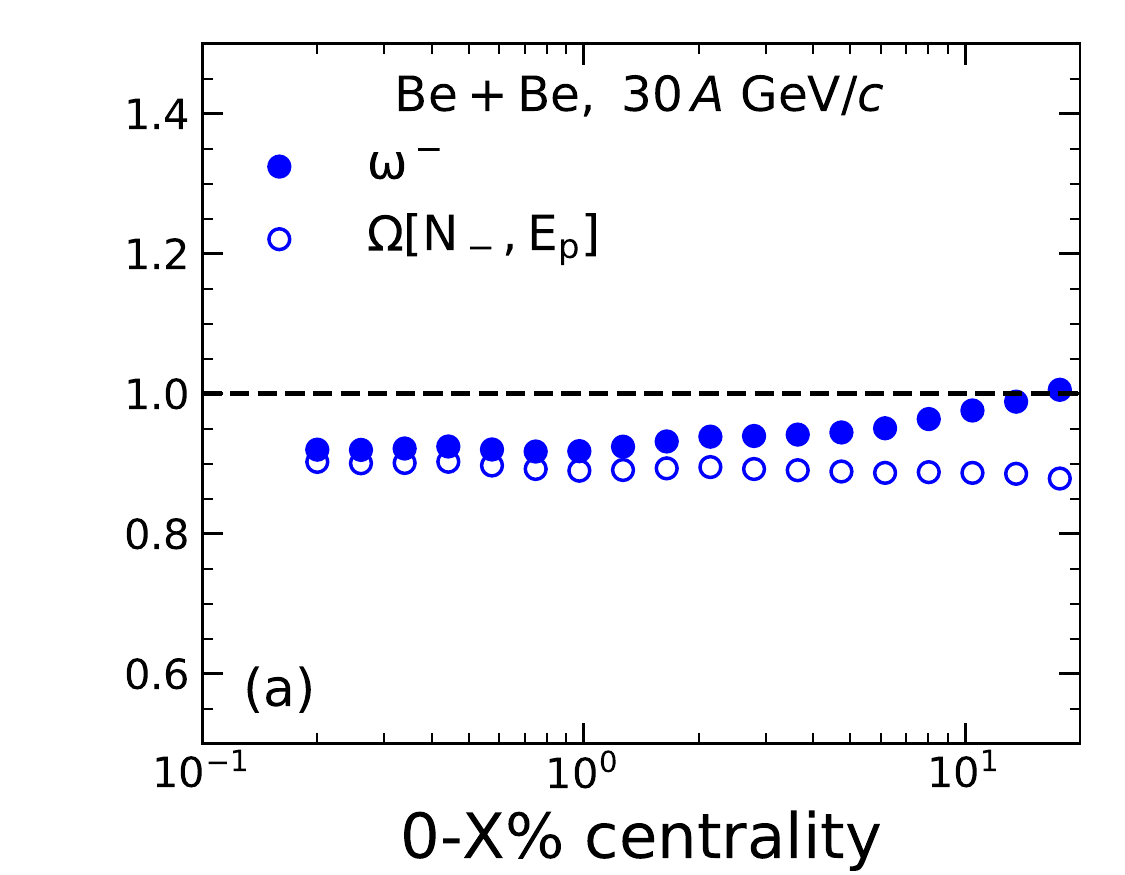}
\includegraphics[width=0.49\textwidth]{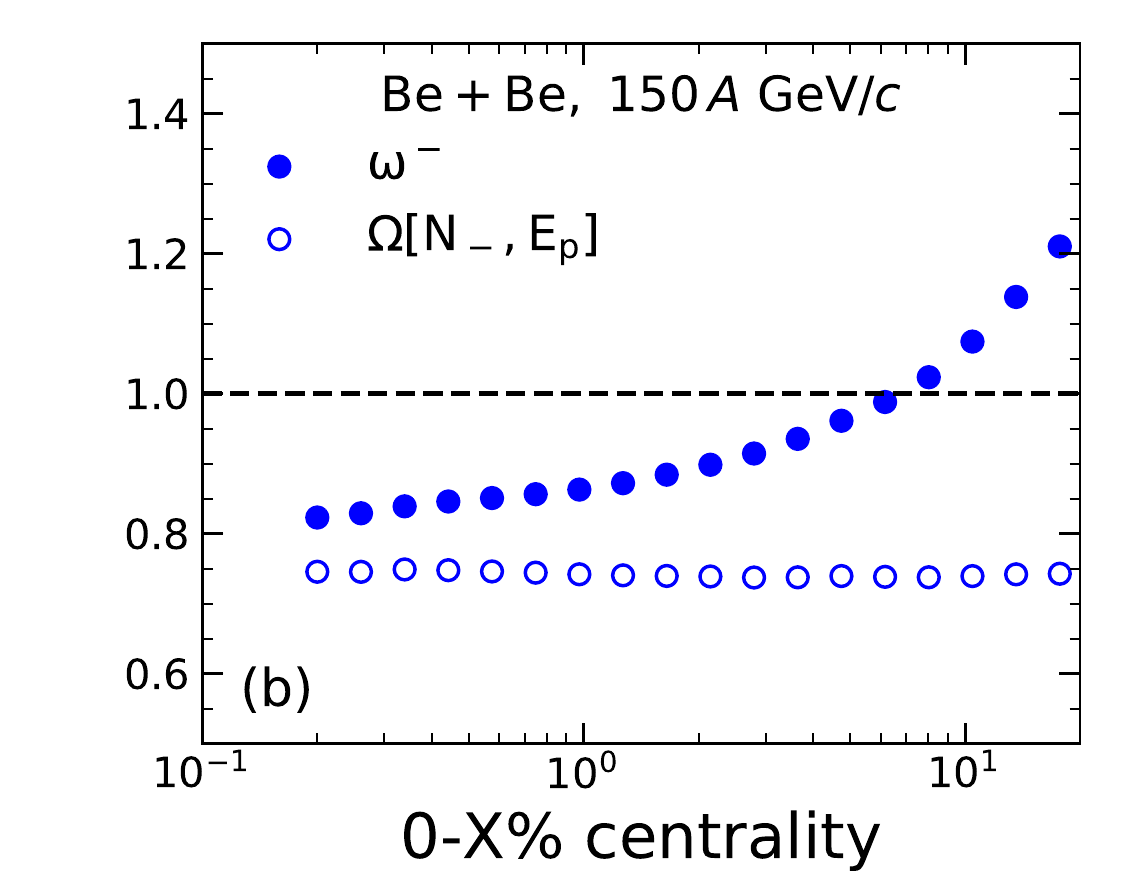}\\
\includegraphics[width=0.49\textwidth]{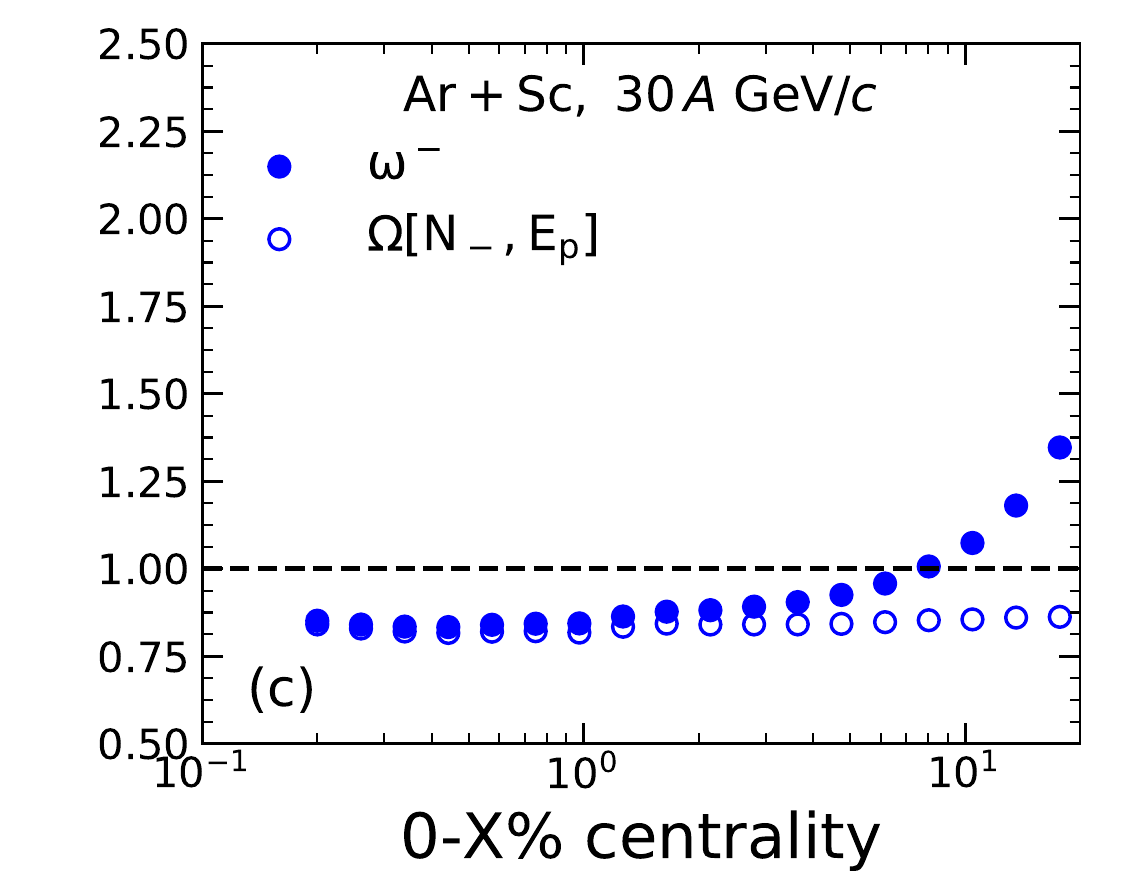}
\includegraphics[width=0.49\textwidth]{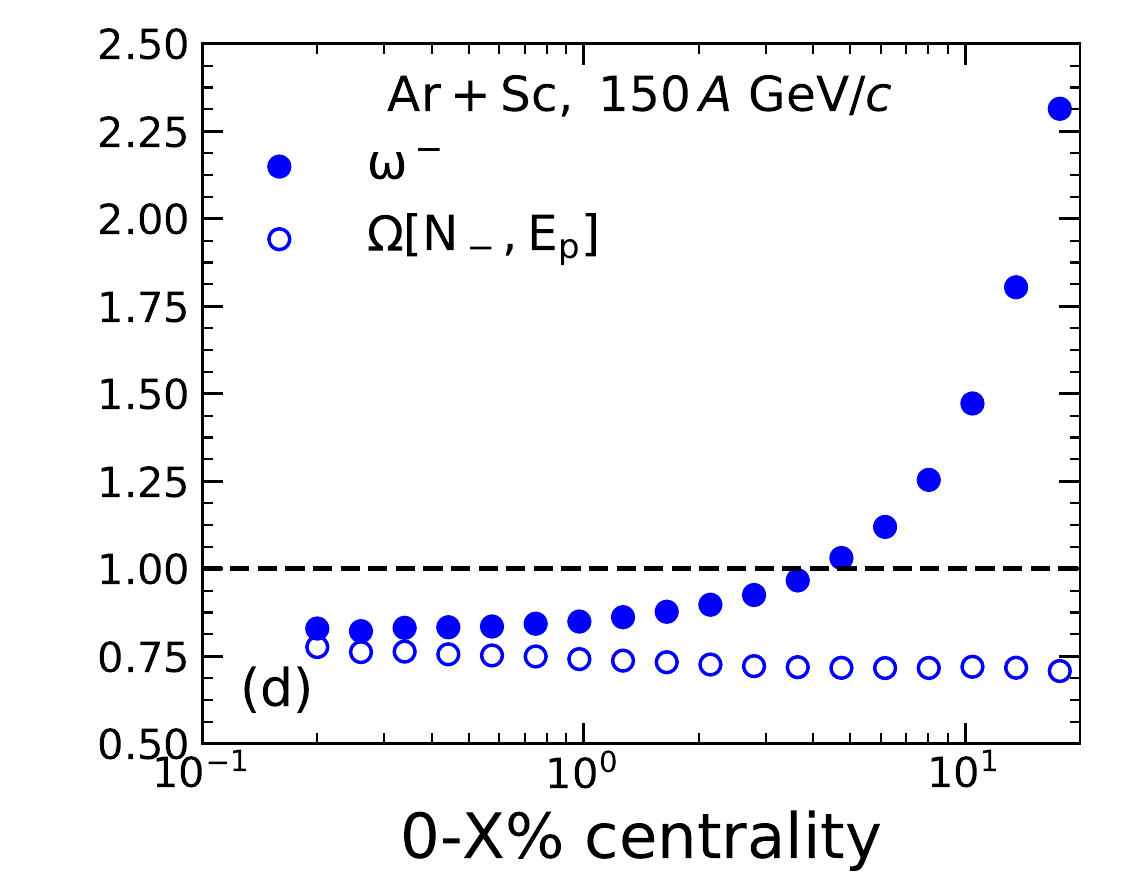}
\caption{The UrQMD results for the scaled variance $\omega^-$ (full symbols) and strongly intensive quantity $\Omega[N_{-},E_{\rm P}]$ (open symbols) as a functions of the centrality for Be+Be (top) and Ar+Sc (bottom) collisions at $p_{\rm lab} = 30~{\rm A\,GeV/c}$ (left) and $p_{\rm lab} = 150~{\rm A\,GeV/c}$ (right).}

\label{fig-omega}
\end{figure}

A behavior of the scaled variance $\omega^-$ as a function of centrality is shown in Fig.~\ref{fig-omega}. For 1\% of most central collision events the values of $\omega^-$ are saturated. An increase of the scaled variance $\omega^-$ in the wider samples of collision events  is seen. It happens because of large fluctuations of the number of nucleon participants. This effect has mostly a simple geometrical origin.

The preferable  centrality class can be also selected by a comparison of $\omega^{-}$ and $\Omega[N_{-},E_{\rm P}]$, where $\Omega[N_{-},E_{\rm P}]$ is a strongly intensive analog of the scaled variance~\cite{Gorenstein:2011vq}:
\eq{
\Omega[N_{-},E_{\rm P}] = \omega^{-} -
(\langle N_-\,E_{\rm P} \rangle -
\langle N_-\rangle\cdot\langle E_{\rm P} \rangle) / \langle E_{\rm P} \rangle\,,
~~~~E_{\rm P}=E_{\rm beam} - E_{\rm F}\,.
}
Figure~\ref{fig-omega} shows that the value of $\Omega[N_{-},E_{\rm P}]$ is not much sensitive to the centrality class. One also observes that the saturation of $\omega^{-}$ does occur in the region of centralities where $\omega^- \cong \Omega[N_-,E_{\rm P}]$.

\subsection{Centrality selection in inelastic p+p reactions}

A model of the independent sources, particularly the wounded nucleon model \cite{Bialas:1976ed}, considers A+A collisions as independent nucleon-nucleon collisions. Under these assumptions the scaled variance $\omega^-$ can be presented as \cite{Bialas:1976ed}
\eq{\label{wnm}
\omega^-_{\rm A+A}~=~\omega^-_{\rm N+N}~+~\frac{1}{2}\,\langle n_-\rangle_{\rm N+N}\,\omega_{\rm part}~,
}
where $\omega^-_{\rm N+N}$ and $\langle n_-\rangle_{\rm N+N}$ are, respectively, the scaled variance and mean multiplicities of negatively charged hadrons in nucleon-nucleon collisions, and $\omega_{\rm part}$  is the scaled variance for the e-by-e fluctuations of the nucleon participants. At $p_{\rm lab}= 150$~GeV/$c$ the approximate relations $\langle n_-\rangle_{\rm N+N}\cong \langle n_-\rangle_{\rm p+p}$ and $\omega^-_{\rm N+N}\cong \omega^-_{\rm p+p}$   appear to be valid~\cite{Konchakovski:2007ss, Konchakovski:2007ah}. Equation (\ref{wnm}) gives then a larger value of $\omega^-$ in any A+A collisions than that in p+p collisions, $\omega^-_{\rm A+A}\ge \omega^-_{\rm p+p}$. To have $\omega^-_{\rm A+A}\cong \omega^-_{\rm p+p}$ one needs $\omega_{\rm part}\cong 0$. In principle, this is possible for a very rigid centrality selection in A+A collisions. However, Eq.~(\ref{wnm}) is evidently in contradiction with $\omega^-_{\rm p+p}> \omega^-_{\rm Pb+Pb}$ seen in the data. 
\begin{figure}[h!]
\centering
\includegraphics[width=0.49\textwidth]{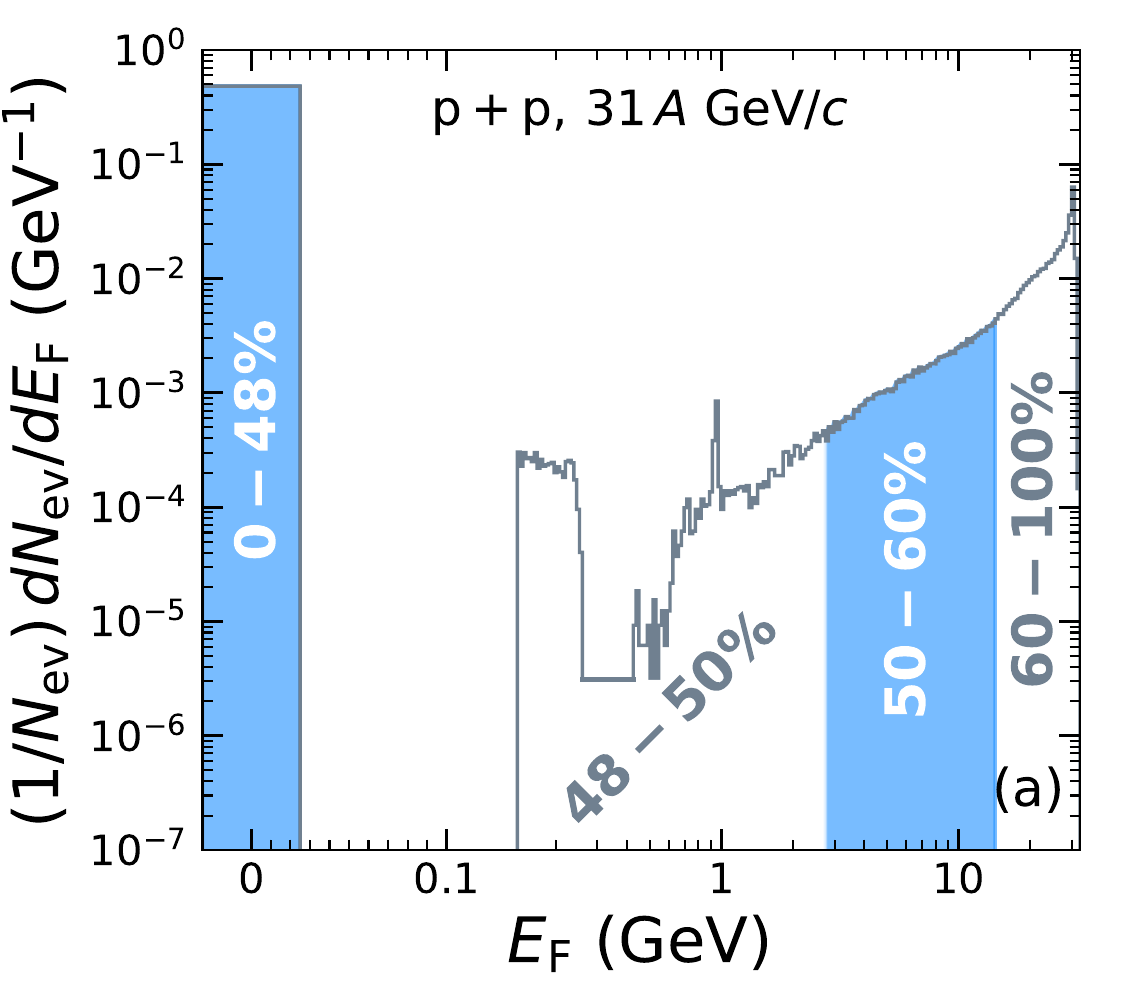}
\includegraphics[width=0.49\textwidth]{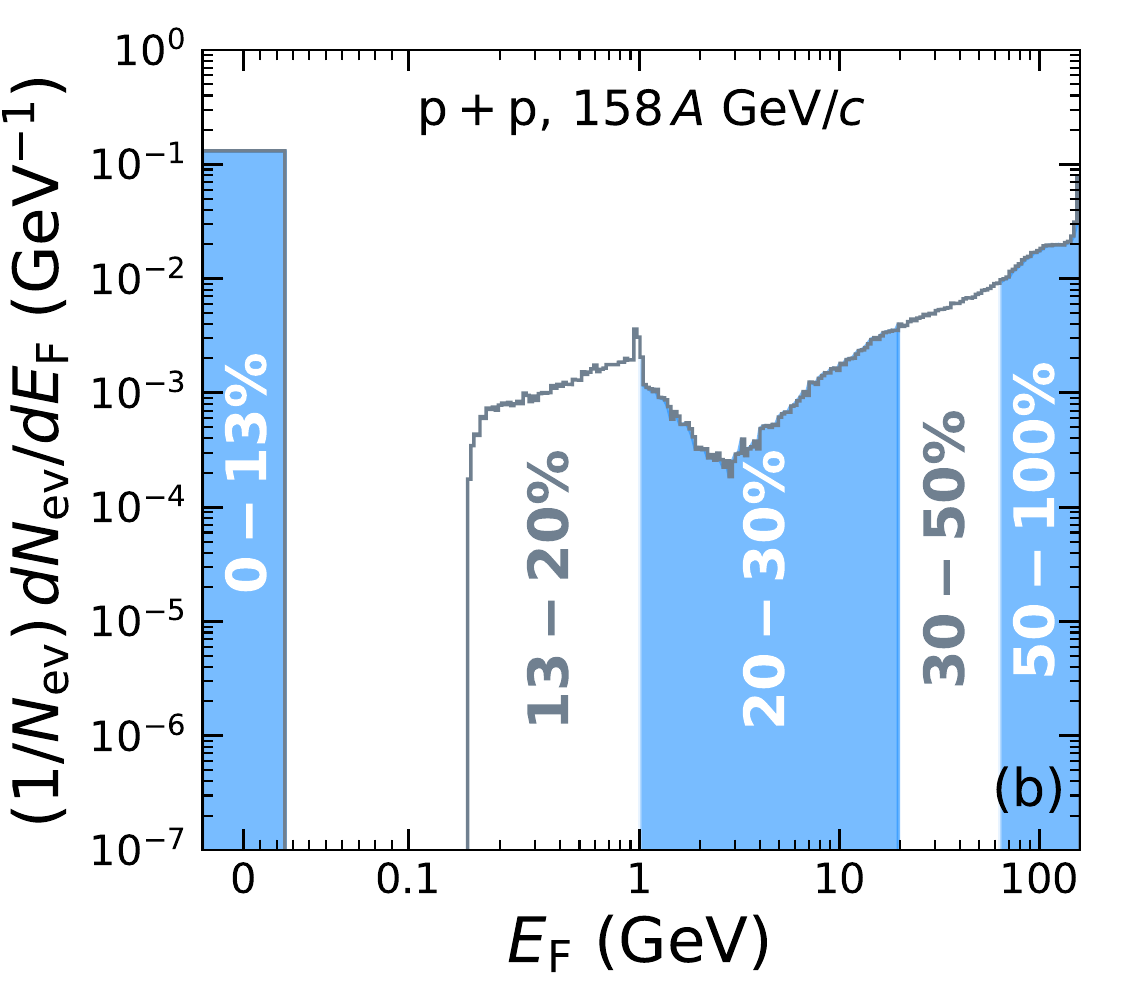}
\includegraphics[width=0.49\textwidth]{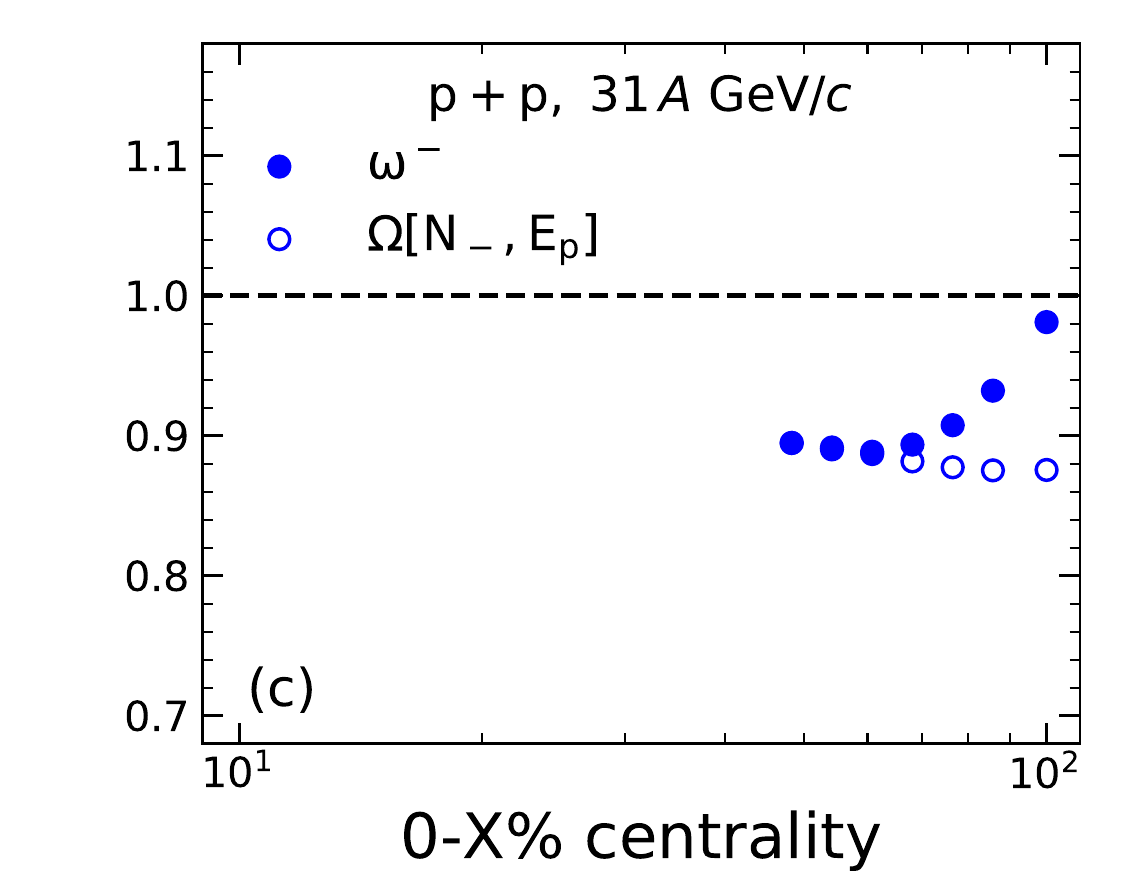}
\includegraphics[width=0.49\textwidth]{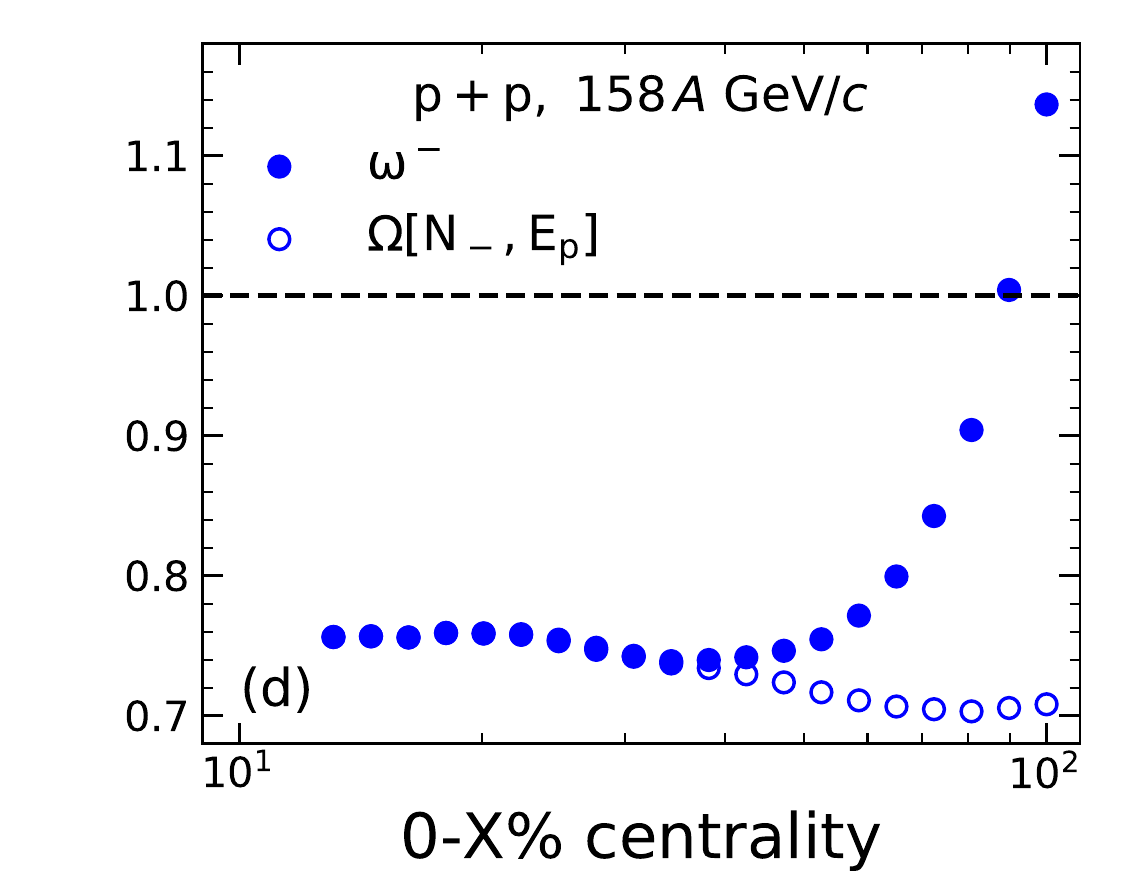}
\caption{The number of events as  functions of the forward energy $E_{\rm F}$ calculated within UrQMD for inelastic p+p collisions at $p_{\rm lab} = 30~{\rm A\,GeV/c}$ (a) and $p_{\rm lab} = 150~{\rm A\,GeV/c}$ (b). The scaled variance $\omega^-$ (full symbols) and strongly intensive measure $\Omega[N_{-},E_{\rm P}]$ (open symbols) as  functions of the centrality for inelastic p+p collisions at $p_{\rm lab} = 30~{\rm A\,GeV/c}$ (c) and $p_{\rm lab} = 150~{\rm A\,GeV/c}$ (d).}
\label{fig-Efwd-pp}
\end{figure}
At the largest SPS collision energies the experimental data suggests $\omega^-_{\rm p+p}>1$,  and  $\omega^-_{\rm A+A}<1$ for most central heavy ion collisions.

What is the origin of $\omega^-_{\rm p+p}>1$ at large collision energies? We argue that the main reason of the large e-by-e fluctuations of hadron multiplicities is an absence of the `centrality selection' in p+p inelastic reactions.
For p+p reactions one considers a sample of all inelastic p+p events. A comparison of the hadron production in {\it most central} A+A events and {\it all} inelastic p+p reactions is not however conclusive. In fact, one expects the strong non-statistical fluctuations in p+p reactions. A presence of these non-statistical fluctuations in p+p reactions is clearly seen at very high collision energies where $\omega^-\gg 1$.

To study this problem we introduce the `centrality selection' for p+p inelastic reactions within the UrQMD simulations. The centrality samples in p+p inelastic reactions will be defined by measuring $E_{\rm F}$ in the same way as in the A+A collisions.  We do not discuss here a physical interpretation of these different centralities in p+p reactions. Because of a lack of the PSD acceptance maps for p+p collisions, we use the Be+Be experimental maps to calculate the energy deposited in the PSD. The UrQMD results for the `centrality samples' in p+p inelastic reactions are shown in Figs.~\ref{fig-Efwd-pp} (a) and (b).
In p+p inelastic reactions there is a non-vanishing  probability that no particle is emitted to the forward energy calorimeter. Among inelastic p+p collisions at $p_{\rm lab}=31~$GeV/$c$ and 158~GeV/$c$ there are, respectively, about  48\% and 13\% of inelastic collision events with $E_{\rm F}=0$. These samples with $E_{\rm F}=0$ will be defined as most central p+p inelastic reactions. Future experimental improvements would make possible a more precise centrality triggering.

The UrQMD values of $\omega^-_{\rm p+p}$ in different centrality classes are shown in Figs.~\ref{fig-Efwd-pp} (c) and (d). At $p_{\rm lab}=158~A~$GeV$/c$ the value of $\omega^-\cong 1.15$ is obtained for all inelastic p+p reactions and $\omega^-\cong 0.76$ for 13\% most central p+p collisions.

\subsection{Comparison with the data}
\label{Sec-Results}

\begin{figure}[h!]
\centering
\includegraphics[width=0.49\textwidth]{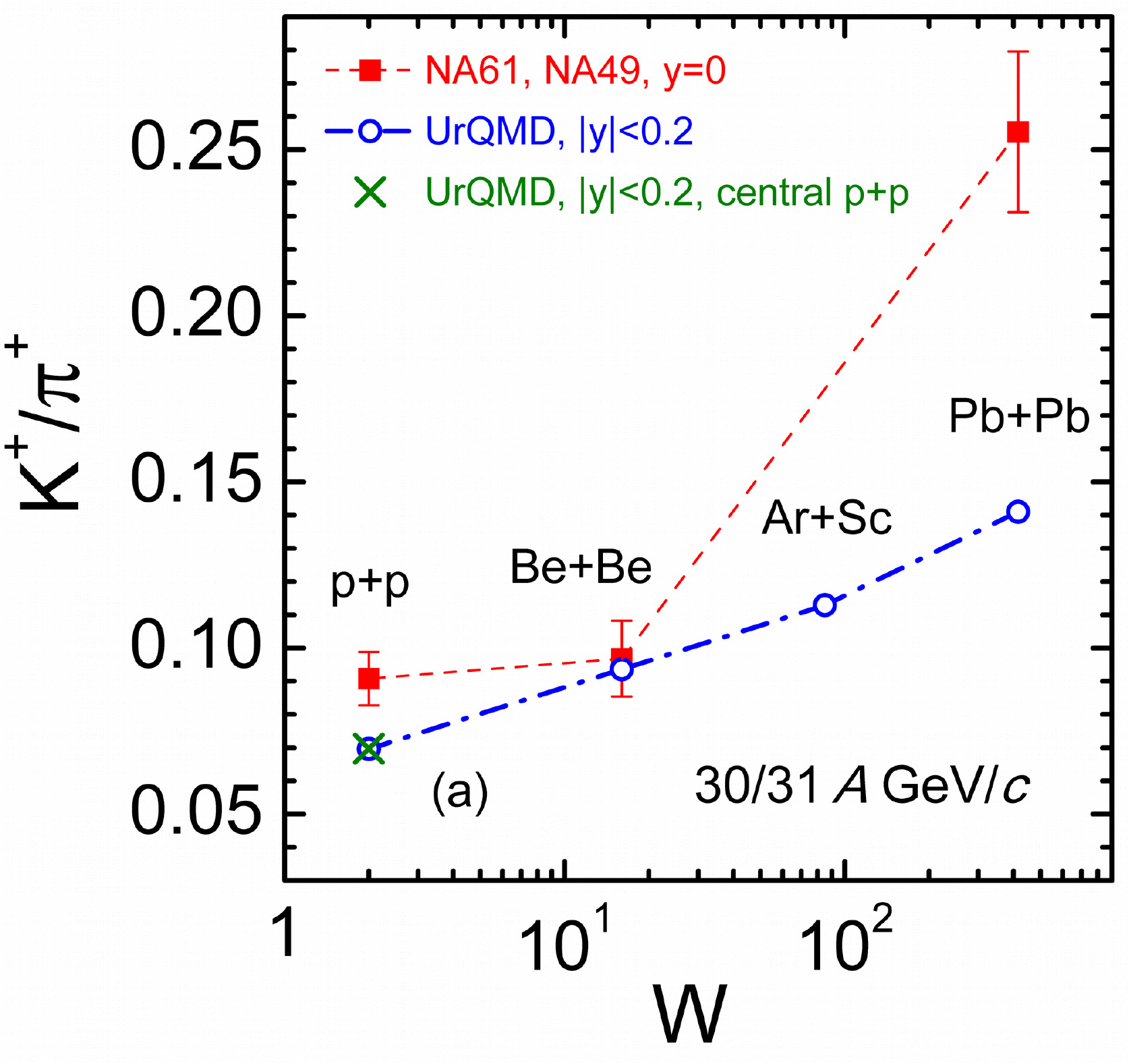}
\includegraphics[width=0.49\textwidth]{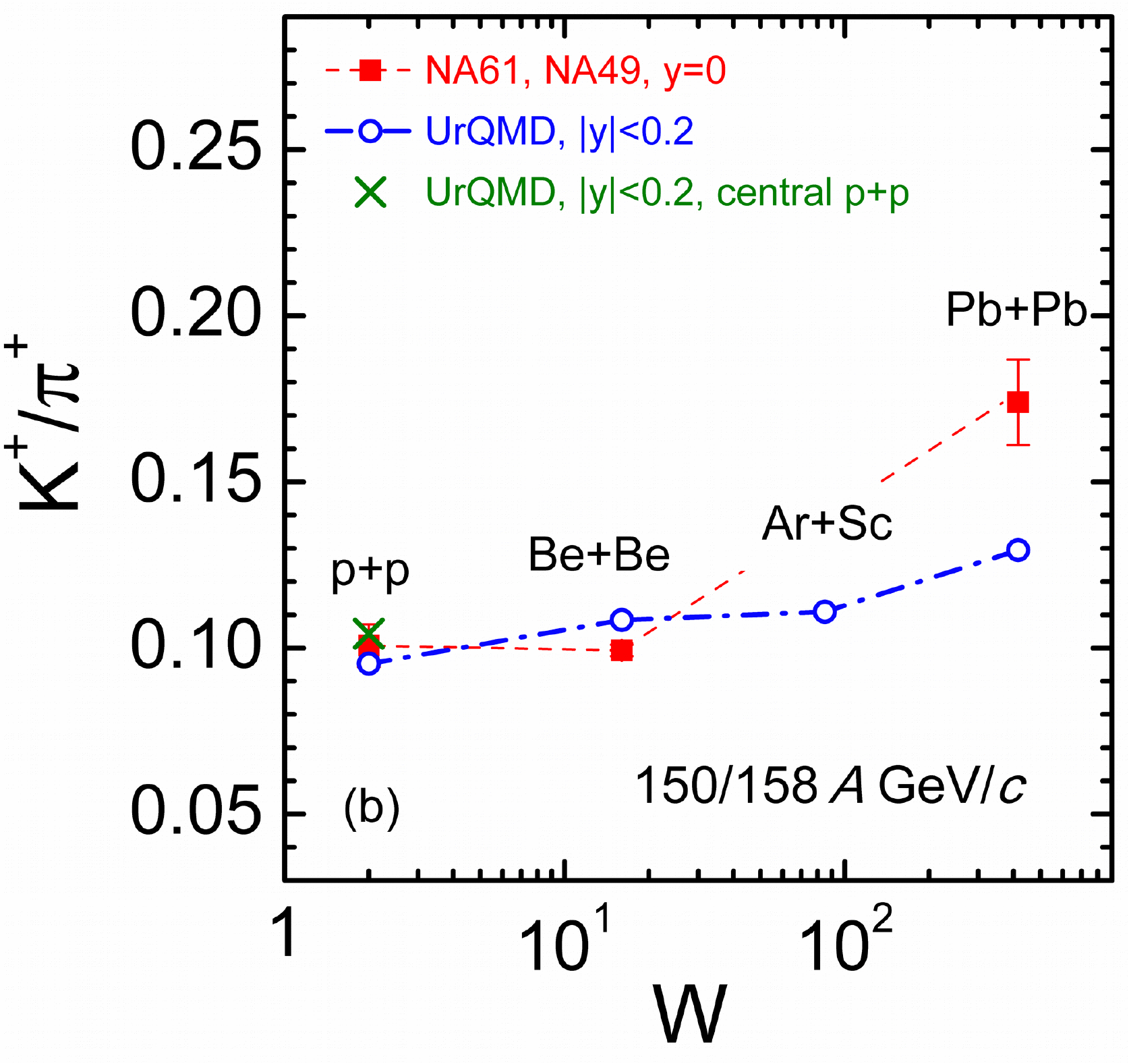}
\\[0.2cm]
\includegraphics[width=0.49\textwidth]{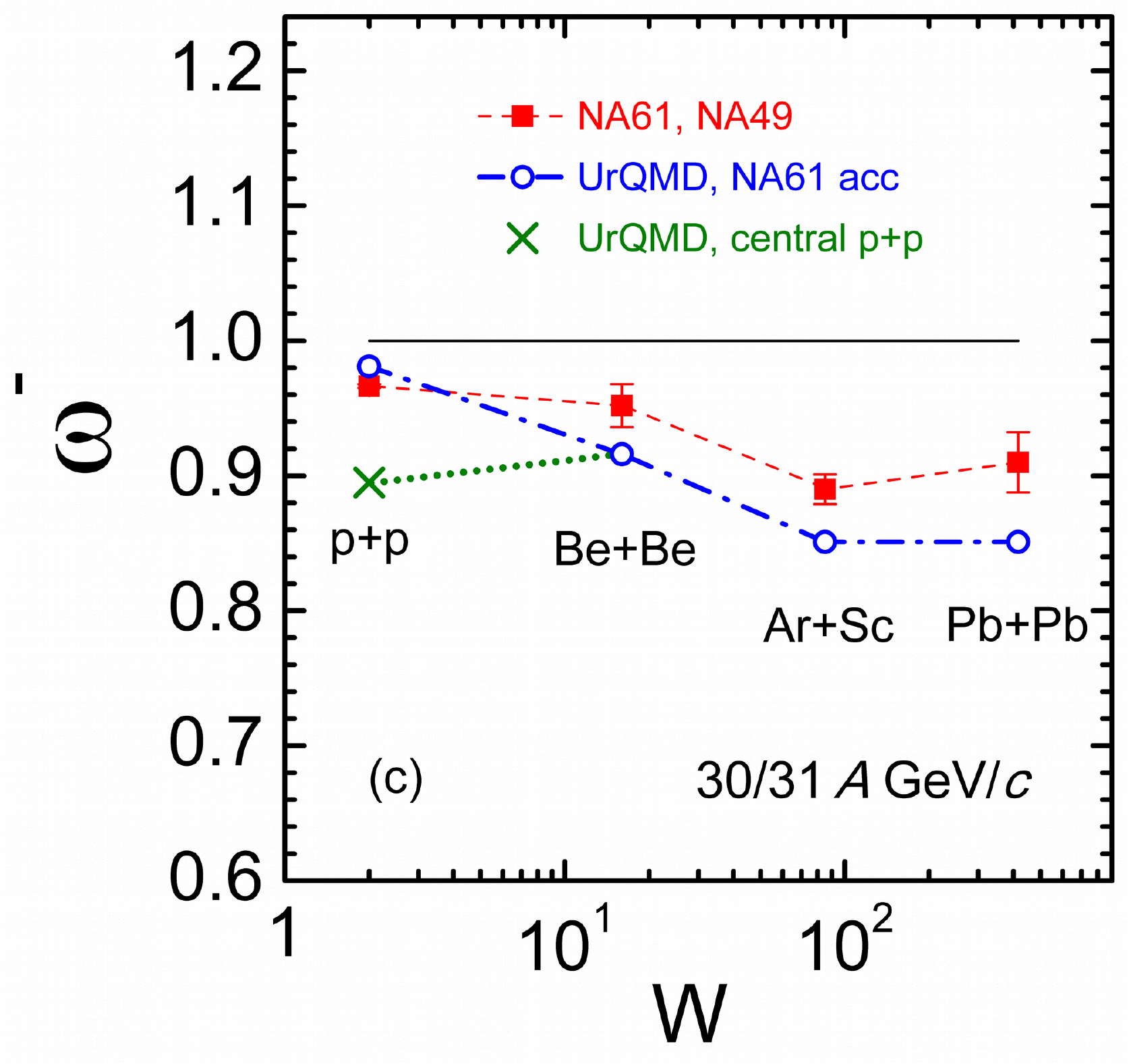}
\includegraphics[width=0.49\textwidth]{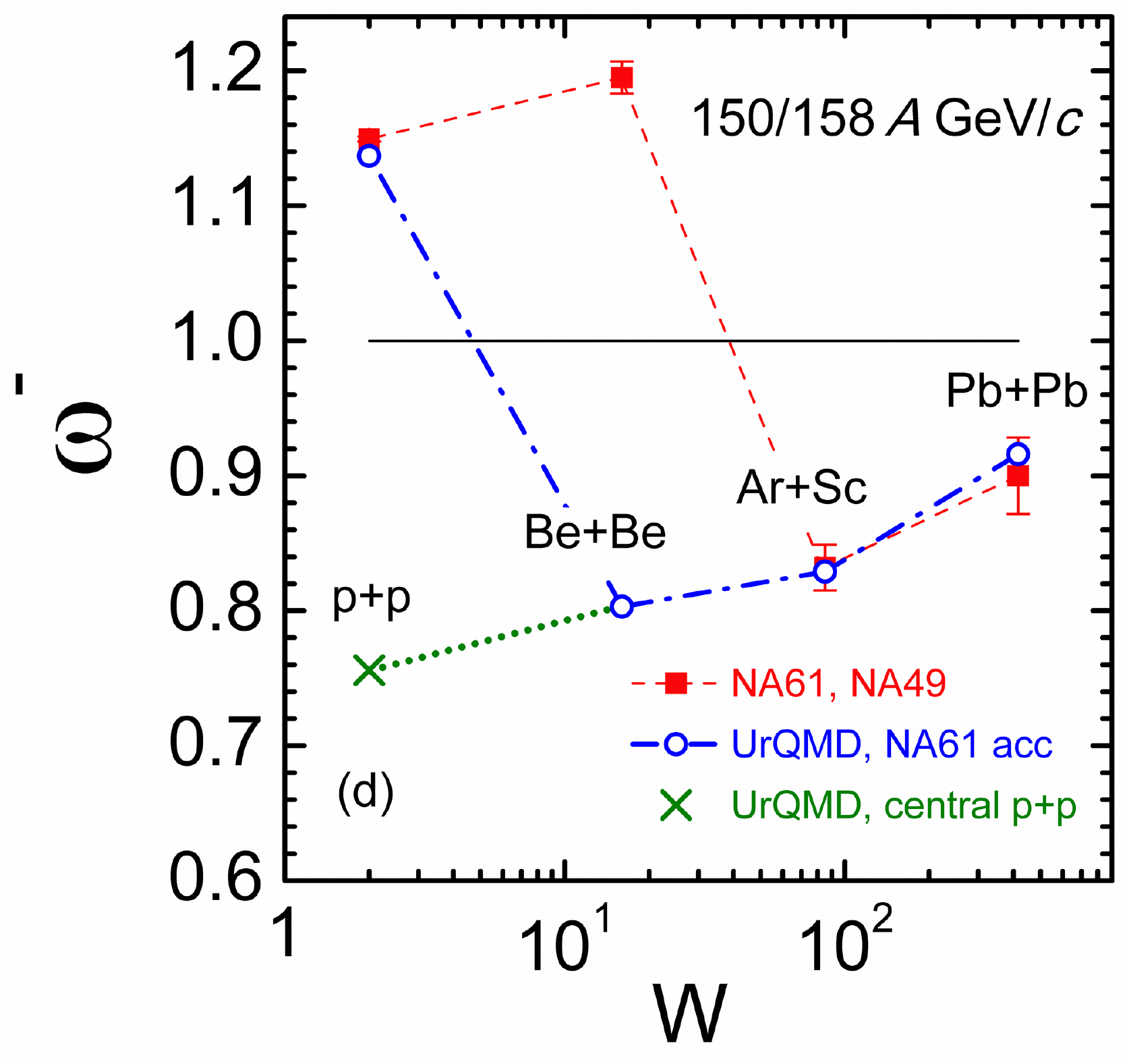}

\caption{The data are the same as in Fig.~\ref{fig-3}. The UrQMD results are shown by open circles.
See text for details. Crosses denote the UrQMD results
for the  most central p+p inelastic collisions.}
\label{fig-HRG-exp-k-pi}
\end{figure}

We present now a comparison of the UrQMD results with the new experimental data
of NA61/SHINE Collaboration and older Pb+Pb data of the  NA49 Collaboration.
The $K^+/\pi^+$ and $\omega^-$ as functions of the system size are
presented in Fig.~\ref{fig-HRG-exp-k-pi}.

As seen from Figs.~\ref{fig-HRG-exp-k-pi} (a) and (b) the UrQMD values of the $K^+/\pi^+$ ratio are systematically smaller than the experimental ones. A main reason of an underestimation of strange hadron production in the UrQMD model is a restriction to the two-body hadron collisions only \cite{Bratkovskaya:2004kv}. In the transport approach which includes the parton degrees of freedom the strange hadron production seems to be in  agreement with the data \cite{Palmese:2016rtq}. On the other hand, a monotonic increase of the $K^+/\pi^+$ ratio with the size of colliding nuclei observed in the data is reproduced by the UrQMD simulations.

As already pointed out, the scaled variance $\omega^-$ is very sensitive to the centrality selection. Figures \ref{fig-HRG-exp-k-pi} (c) and (d)  present a comparison of the UrQMD results with the available data for $\omega^-$. The UrQMD demonstrates a good agreement with the $\omega^-$ data for Ar+Sc and Pb+Pb central collisions, and for all inelastic p+p collisions. There is only one reaction which is out of the description -- Be+Be at 150$A$ GeV/$c$. Note that this data point has still a preliminary status, and additional experimental checks are in progress~\cite{GazdPriv}.

For the UrQMD simulations of p+p reactions we also present the results for `most central' p+p events calculated in the samples with $E_{\rm F}=0$. These results are presented by crosses.
The centrality selection performed for p+p reactions within the UrQMD simulations leads to essentially smaller values of $\omega^-_{\rm p+p}$.

%
\section{Summary}
\label{Sec-Summary}
The $K^+/\pi^+$ ratio and the scaled variance for negatively charged particles $\omega^-$ are calculated within the HRG and UrQMD models in p+p and A+A collisions at the SPS energy range.
The HRG calculations are done within the CE with the three conserved charges -- baryon number, electric charge, and strangeness. A qualitative agreement with the new data of the NA61/SHINE Collaboration -- a monotonous increase of the $K^+/\pi^+$ ratio with a size of colliding nuclei -- is observed. The strangeness suppression factor $\gamma_S < 1$ reflecting an incomplete strangeness equilibration leads to improvement of the quantitative agreement of the CE HRG results with the $K^+/\pi^+$ data.

The data on  $\omega^-< 1$ in Ar+Sc and Pb+Pb most central collisions can be described by the CE HRG. However, for the two smallest systems, p+p and Be+Be, at the highest SPS energy, the CE HRG results are in contradiction with the large experimental values of $\omega^- > 1$.

The UrQMD calculations are done by applying the acceptance cuts and event selection procedure used by the NA61/SHINE Collaboration. Particularly, the centrality selection is done using the forward  energy $E_{\rm F}$ deposited in the PSD.
The UrQMD explains the qualitative trends of the $K^+/\pi^+$ ratio.
A comparison of the UrQMD results with the data on $\omega^-$
looks well except for only one point -- Be+Be at $150A$~GeV$/c$.

Two effects -- statistical and dynamical -- are identified in the e-by-e fluctuations of hadron production at the SPS energies. Statistical effects are clearly seen in the both CE HRG and UrQMD calculations. This is first of all a suppression of the particle number fluctuations because of the global charge conservation. Contrary to  naive expectations these suppression effects are stronger for large collective systems. On the other hand, one also sees a presence of the dynamical fluctuations. They are most clearly pronounced in the sample of all inelastic p+p reactions at high collision energy. The experimental value of the scaled variance $\omega^-$ becomes larger than unity. This can not be explained within the CE HRG model. The UrQMD model takes the p+p data, both hadron yields and their e-by-e fluctuations, as the input to describe A+A collisions. Thus, the model takes into account these dynamical effects in p+p reactions.  The only reaction for which the value of $\omega^-$ is not described by the UrQMD simulation is Be+Be at 150~$A$~GeV/$c$.  Note that the UrQMD results for $\omega^-$ in this reaction appear to be extremely sensitive to the exact centrality selection procedure.

To clarify the dynamical features of the  e-by-e fluctuations we propose  to implement the centrality selection procedure in p+p reactions.  A comparison of p+p and A+A collisions should be done with the appropriate centrality selection procedures in both reactions.
%


\acknowledgments
The authors thank E.~Bratkovskaya, M.~Gazdzicki, M.~Kuich, O.~Linnyk, A.~Seryakov, and J.~Steinheimer for fruitful discussions. A.M. appreciates useful discussions with the participants of the ``NA61/SHINE Collaboration Meeting'', September 10-14, CERN, Switzerland. The work of M.I.G. is supported by the Alexander von Humboldt Foundation and by the Goal-Oriented Program of Cooperation between CERN and National Academy of Science of Ukraine "Nuclear Matter under Extreme Conditions" (agreement CC/1-2018).

\bibliography{fireball}

\end{document}